\documentclass{aa}

\usepackage{graphicx}
\usepackage{txfonts}

\usepackage{hyperref}
\hypersetup{
    colorlinks,
    linkcolor=blue,
    citecolor=blue,
    urlcolor=blue
}

\usepackage{booktabs}       % professional-quality tables
\newcommand{\crbox}{\textsc{SLOW-CR3072}$^3$}
\newcommand{\comazoom}{\textsc{Coma}}
\newcommand{\comalow}{\textsc{Coma}-$D_\mathrm{pp}$}
\newcommand{\bsim}{$B_\mathrm{sim}$}
\newcommand{\bff}{$B_\mathrm{ff}$}
\newcommand{\bbeta}{$B_{\beta}$}
\newcommand{\bturb}{$B_{\mathcal{F}}$}
\newcommand{\bdynl}{$B_{\mathrm{dyn},\downarrow}$}
\newcommand{\bdynh}{$B_{\mathrm{dyn},\uparrow}$}

\graphicspath{{Figures/}}
\begin{document}

\title{Simulating the LOcal Web (SLOW)\\ III. Synchrotron emission from the local cosmic web}

   %\subtitle{Synchrotron Emission from the Local Cosmic Web}

   \author{Ludwig M. B\"oss\thanks{\email{lboess@usm.lmu.de}}\fnmsep
          \inst{1}
          \and
          Klaus Dolag\inst{1,2}
          \and
          Ulrich P. Steinwandel\inst{3}
          \and 
          Elena Hernández-Martínez\inst{1}
          \and\\
          Ildar Khabibullin\inst{1,2}
          \and
          Benjamin Seidel\inst{1}
          \and
          Jenny G. Sorce\inst{4,5,6}
          }

   \institute{Universit\"ats-Sternwarte, Fakult\"at f\"ur Physik, Ludwig-Maximilians-Universit\"at M\"unchen, Scheinerstr.1, 81679 M\"unchen, Germany
         \and
             Max Planck Institute for Astrophysics, Karl-Schwarzschild-Str. 1, D-85741 Garching, Germany
\and
Center for Computational Astrophysics, Flatiron Institute, 162 5th Avenue, New York, NY 10010, USA
\and
Univ. Lille, CNRS, Centrale Lille, UMR 9189 CRIStAL, F-59000 Lille, France
\and
Universit\'e Paris-Saclay, CNRS, Institut d'Astrophysique Spatiale, 91405, Orsay, France
\and
Leibniz-Institut f\"{u}r Astrophysik (AIP), An der Sternwarte 16, D-14482 Potsdam, Germany
             }

   \date{Received XXXX; accepted YYYY}

% \abstract{}{}{}{}{} 
% 5 {} token are mandatory
  \abstract
  % context heading (optional)
  % {} leave it empty if necessary  
   {}
  % aims heading (mandatory)
   {Detecting diffuse synchrotron emission from the cosmic web is still a challenge for current radio telescopes. We aim to make predictions about the detectability of cosmic web filaments from simulations.}
  % methods heading (mandatory)
   {We present the first cosmological magnetohydrodynamic simulation of a 500 $h^{-1} c$Mpc volume with an on-the-fly spectral cosmic ray (CR) model. This allows us to follow the evolution of populations of CR electrons and protons within every resolution element of the simulation. We modeled CR injection at shocks, while accounting for adiabatic changes to the CR population and high-energy-loss processes of electrons. The synchrotron emission was then calculated from the aged electron population, using the simulated magnetic field, as well as different models for the origin and amplification of magnetic fields.
   We used constrained initial conditions, which closely resemble the local Universe, and compared the results of the cosmological volume to a zoom-in simulation of the Coma cluster, to study the impact of resolution and turbulent reacceleration of CRs on the results.}
  % results heading (mandatory)
   {We find a consistent injection of CRs at accretion shocks onto cosmic web filaments and galaxy clusters. This leads to diffuse emission from filaments of the order $S_\nu \approx 0.1 \: \mu$Jy beam$^{-1}$ for a potential LOFAR observation at 144~MHz, when assuming the most optimistic magnetic field model.
   The flux can be increased by up to two orders of magnitude for different choices of CR injection parameters.
   This can bring the flux within a factor of ten of the current limits for direct detection.
   We find a spectral index of the simulated synchrotron emission from filaments of $\alpha \approx$ -1.0 -- -1.5 in the LOFAR band.}
   {}

   \maketitle

%%%%%%%%%%%%%%%%%%%%%%%%%%%%%%%%%%%%%%%%%%%%%%%%%%

%%%%%%%%%%%%%%%%% BODY OF PAPER %%%%%%%%%%%%%%%%%%

% introduction
\section{Introduction}

From observations of galaxy clusters, we can assume an efficient acceleration of cosmic ray (CR) electrons (CRes), which provide powerful tracers of the intra-cluster medium's magnetic field such as radio relics and radio halos \citep[see][for a review]{Weeren2019}.
The recent discovery of “radio mega-halos” \citep[][]{Cuciti2022} and “radio bridges” \citep[e.g.,][]{Govoni2019, Botteon2020a, Bonafede2021, Venturi2022,  Radiconi2022} also indicates that there must be a substantial population of relativistic electrons even further from the cluster center than previously expected.
The goal of this work is to explore if this is also true in cosmic web filaments.\\
The existence of a volume-filling, relativistic electron population poses a number of theoretical problems for the acceleration of electrons and their stabilization against energy losses, as well as magnetic field strength in these regimes.
The canonical process of CR acceleration in galaxy clusters is diffusive shock acceleration (DSA) \citep[see][for a review]{Drury1983}, where particles are accelerated from the thermal pool by scattering off magnetohydrodynamic (MHD) turbulence up and downstream of shock fronts, gaining energy at every crossing from upstream to downstream.
This naturally produces a power-law distribution of CRs in momentum space, consistent with the power-law slope of the synchrotron spectrum of radio relics.
Hybrid- and particle-in-cell (PIC) simulations of CR acceleration find that the acceleration efficiency at the shock depends strongly on the sonic Mach number \citep[see e.g., studies by][]{Kang2013, Caprioli2014, Ryu2019}, which would indicate a highly efficient acceleration of CRs at the periphery of the cluster; for example, at accretion shocks around galaxy clusters and cosmic filaments where high Mach number shocks are found in simulations \citep[e.g.,][]{Ryu2003, Pfrommer2006, Vazza2009, Vazza2011, Schaal2015, Banfi2020, Ha2023}.
However, detailed studies on the efficiency of CRe acceleration in high-$\beta$ (with $\beta = \frac{P_\mathrm{th}}{P_\mathrm{B}} \sim 50 - 10^3$) plasmas are typically performed for high-temperature, low-Mach-number shocks \citep[e.g.,][]{Guo2014, Kang2019, Kobzar2021, Ha2021} to understand the discrepancy between the predicted electron acceleration efficiency and radio brightness of radio relics \citep[see][for a detailed discussion]{Botteon2020}.
\citet{Ha2023} recently performed PIC simulations with a high-$\beta$, low-temperature ($T = 10^4 $ K), and high-sonic-Mach-number ($\mathcal{M}_s = 25 - 100$) setup to study the acceleration efficiency of CRe in accretion shocks.
Applying their model to post-process outputs from a cosmological simulation, they find emission of the order $S_\nu \sim 1 - 10$~mJy~beam$^{-1}$ for accretion shocks around a Coma-like cluster, which is in agreement with emission by an accretion shock reported by \citet{Bonafede2022}.

The second ingredient to potential synchrotron emission in filaments is the magnetic field strength in these environments.
These field strengths can be estimated from rotation measurements, where the intrinsic polarization angle of a strongly polarized source, like synchrotron emission by electrons accelerated in an active galactic nucleus (AGN) jet, is rotated by the magnetic field along the line of sight to the observer.
These observations indicate magnetic field strengths of 10-100 nG \citep[e.g.,][]{Vernstrom2019, Carretti2022, OSullivan2019, OSullivan2020, OSullivan2023}.
Alternatively, the stacking of cluster pairs and filaments can provide upper limits on the synchrotron emission and with that the superposition of CRe population and magnetic field strength \citep[e.g.,][]{Brown2017, Vernstrom2017, Vernstrom2021, Locatelli2021, Hoang2023}.
Direct observations of radio emission from cosmic web filaments could help to disentangle this superposition, and therefore give an insight into the CRe acceleration, as well as the magnetic field strength and structure in cosmic web filaments. Several numerical studies have made predictions about such observations by taking into account different models for magnetic fields and CRe components and applying them to large-scale simulations.

\citet{Vazza2015} used MHD simulations of a (50 Mpc)$^3$ box \citep[][]{Vazza2014} and applied shock detection and the CR acceleration model of \citet{Hoeft2007} to study potential synchrotron emission by shock-accelerated electrons in accretion shocks and internal shocks in cosmic web filaments.
They varied their magnetic field strength by rescaling their simulated magnetic field with a density model and found in the high amplification model a diffuse emission from filaments on the order of $\sim \mu$Jy beam$^{-1}$.

\citet{Brown2017} used a different approach by modeling CR electrons as secondaries of shock-accelerated protons.
These protons are expected to live for longer than one Hubble time in the low-density cosmic web filaments and can scatter with thermal protons into charged pions, which decay into electrons and positrons \citep[e.g.,][]{Blasi2007}.
These secondary models require a substantial primary proton population to produce considerable secondary electrons, which is in conflict with the current non-detection of $\gamma$-ray photons that should also originate from this process \citep[see e.g.,][and references therein]{Wittor2019}.
\citet{Brown2017} took the model by \citet{Dolag2004}, which constructs radio power based on secondaries from a volume-filling proton population with a proton to thermal pressure ratio of $X_\mathrm{cr} = 0.01$, and applied it to a constrained simulation resembling the local cosmic web \citep[][]{Dolag2004, Dolag2005}{}{}.
This allowed them to use S-PASS data to constrain the diffuse flux from synchrotron emission in cosmic web filaments to $I_{1.4 \mathrm{GHz}} < 0.073
\mu$Jy arcsec$^{-2}$, which is just below the current detection limit of, for example, the EMU survey \citep[][]{Norris2011}{}{}.

\citet{Oei2022} used a (100 Mpc)$^3$ cosmological MHD simulation by \citet{Vazza2019} to obtain synchrotron emission by accretion and merger shocks, again following the \citet{Hoeft2007} model.
They constructed a specific intensity function of the synchrotron cosmic web, by ray-tracing through their simulation box, while taking into account the cosmological distance effects of the emitted radiation.
With this, they find specific intensities of the order $I_\nu \xi^{-1} \sim 0.1$Jy deg$^{-2}$ at $\nu = 150$ MHz, with $\xi$ being the poorly constrained efficiency parameter of electron acceleration.

Here, we take a different approach to a prediction of synchrotron emission from the cosmic web, by injecting and evolving a population of CRe at shocks and studying the magnetic field strengths required to obtain synchrotron emission compatible with current observational limits.
In this work, we study potential emission from CRs accelerated at shocks in merger and accretion shocks of galaxy clusters and accretion shocks around cosmic web filaments.
For this, we employed the first realization of a cosmological MHD simulation of a (500 $h^{-1}c$Mpc$)^3$ box with an on-the-fly Fokker-Planck solver to model CR proton and electron evolution.
Here, we only focus on the electron populations, leaving the study of protons to future work.
This paper is structured as follows.
In Sect.~\ref{sec:methods}, we describe the simulation and the code configuration employed for this work.
In Sect.~\ref{sec:allsky}, we study full sky projections of the injected CR electrons and their synchrotron emission under different magnetic field models.
In Sect.~\ref{sec:coma-filament}, we focus on the analysis of nonthermal emission from our replica of the Coma cluster and its surrounding filament structure.
Section~\ref{sec:box} extends this analysis to the full simulation domain to study the emissivity of all cosmic web filaments.
In Sect.~\ref{sec:discussion}, we compare our results to the post-processing model of \citet{Hoeft2007} and discuss the impact of free parameters in our CR model on our results and their observational prospects.
Finally, Sect.~\ref{sec:conclusions} contains our conclusions and a summary of our work.

% methods
\section{Methods}
\label{sec:methods}
\subsection{Initial conditions}
\begin{table*}
    \centering
    \caption{Presented simulations.}
    \begin{tabular}{|l|c|c|c|c|c|}
    %\toprule
        Name & $M_\mathrm{gas}$ [$M_\odot$]& $M_\mathrm{DM}$ [$M_\odot$] & $\epsilon$ [$h^{-1} c$kpc] & $h_\mathrm{sml, min}$ [kpc] & $D_\mathrm{0, max}$ [s$^{-1}$]\\
        \midrule
        \crbox & $8.5 \times 10^7$ & $4.6 \times 10^8$ & 2.7 & 6.24 & 0 \\
        \comazoom & $1.0 \times 10^7$ & $5.7 \times 10^7$ & 1.0 & 3.6 & 0 \\
        \comalow & $1.0 \times 10^7$ & $5.7 \times 10^7$ & 1.0 & 3.7 & $10^{-18}$
     %  \bottomrule
    \end{tabular}
    \tablefoot{From left to right, we list the name, mass of the gas particles, mass of the DM particles, gravitational softening, the maximum resolution at $z=0$, and the maximum allowed value for $D_0$.}
    \label{tab:simulations}
\end{table*}
The initial conditions of the simulations have been described extensively in \citet{Sorce2018, Dolag2023}, so we limit the description here to a short overview.

We set up constrained initial conditions based on velocity field reconstructions in the local Universe \citep[][and references therein]{Carlesi2016, Sorce2018}.
In this method, a Wiener filter algorithm \citep[][]{Zaroubi1995, Zaroubi1998} is applied to reconstruct the 3D peculiar velocities of galaxies from the \textsc{CosmicFlows-2} distance modulus survey \citep[][]{Tully2013}.
This survey undergoes several treatments upstream and downstream of this reconstruction \citep[][]{Doumler2013a, Doumler2013b, Doumler2013, Sorce2014, Sorce2015, Sorce2017, SorceTempel2017, SorceTempel2018}.
Subsequently, using the constrained realization algorithm \citep[][]{Hoffman1991}, it can be used to construct a density field of the local Universe at an initial redshift; in other words, the constrained initial conditions.
The resolution of the density field is then increased to reach our study requirement using the \textsc{Ginnungagap} software\footnote{https://code.google.com/p/ginnungagap/}.

These initial conditions have been used, for instance, to study individual clusters such as Virgo \citep[][]{Sorce2016b, Olchanski2018, Sorce2019}, as well as the large-scale environment around the Local Group \citep[][]{Carlesi2016, Carlesi2017} and other clusters \citep[][]{Sorce2023}, in particular the Coma cluster \citep[][]{Malavasi2023}.
Previous work on ultrahigh-energy CRs has shown how these constrained initial conditions help to provide predictions for observations \citep[][]{Hackstein2018}.

In this work, we study the \crbox~simulation, which consists of $2\times3072^3$ particles in a volume of $V = (500 \: h^{-1} \: c$ Mpc$)^3$.
This leads to a mass resolution of $M_\mathrm{gas} \approx 8.5 \times 10^7 M_\odot$ and $M_\mathrm{DM} \approx 4.6 \times 10^8 M_\odot$ for gas and DM particles, respectively.
The maximum resolution of gas particles in our simulation corresponds to $h_\mathrm{sml, min} \approx 6.24$ kpc at $z=0$.

To test the impact of resolution on our results, we also performed zoom-in simulations of the Coma cluster at eight times the resolution of the cosmological box.
To create such zoom-in initial conditions, a dark-matter-only simulation has been extended far into the future, for example up to an expansion factor of $a\approx1000$, to identify the gravitationally bound region of the supercluster that the target cluster is a part of.
Then, the spatial region in the simulation that contains all particles bound to the Coma supercluster was identified.
This region was then traced back to the corresponding Lagrangian volume in the initial conditions, which was then sampled with the particles taken from much higher-resolution versions of the box; in this case, the one sampled with $6144^3$ particles in total. In order to smooth the resolution difference at the boundaries while minimizing the total particle number, boundary layers were added to the high-resolution region, using the lower-resolution versions of the full box in descending order of particle number ($3072^3$, $1536^3$, $768^3$, and $384^3$ for the remainder of the box).
The refinement of the initial conditions leads to a mass resolution for the high-resolution region of $M_\mathrm{gas} \approx 1 \times 10^7 M_\odot$ and $M_\mathrm{DM} \approx 5.7 \times 10^7 M_\odot$ for gas and DM particles, respectively.
Here, we reached a maximum resolution of gas particles of $h_\mathrm{sml, min} \approx 3.7$~kpc.
We list all simulations discussed in this work in Table \ref{tab:simulations}.

All simulations were run using a Planck cosmology \citep[][]{Planck2014} with matter densities of $\Omega_\mathrm{m} = 0.307$ and $\Omega_\mathrm{baryon} = 0.048$, a cosmological constant of $\Omega_\Lambda = 0.692$, and the Hubble parameter $H_0 = 67.77$ km s$^{-1}$ Mpc$^{-1}$.
\subsection{Simulation code}
We employed \textsc{OpenGadget3} \citep[][]{Groth2023}, an advanced version of the cosmological \textsc{Tree-SPH} code \textsc{Gadget2} \citep{Springel2005}.
\textsc{OpenGadget3} uses a Barnes and Hut tree \citep[][]{Barnes1986} to solve the gravity at short range and a particle-mesh (PM) grid for long-range forces.
For the zoom-in simulations used in this work, the PM grid was split to add a higher-resolution grid to the central region to save computing time.

All simulations presented in this work are non-radiative CR-MHD simulations.
\textsc{OpenGadget3} uses an updated smoothed particle hydrodynamics (SPH) implementation \citep[][]{Beck2016} with a spatially and time-dependent high-resolution shock capturing scheme \citep[][]{Dolag2005b, Cullen2010}.
The simulations presented here were run with a Wendland $C_4$ kernel with 200 neighbors and bias correction \citep[as introduced in][]{Dehnen2012} to ensure stabilization against the tensile (pairing) instability that arises for kernels that do not have a positive definite Fourier transformation.
Our SPH implementation was further stabilized by a high-resolution maximum entropy scheme, which in this work was realized by physical rather than artificial thermal conduction.

The MHD solver was presented in \citet{Dolag2009} and extended in \citet{Bonafede2011} to include nonideal MHD in the form of magnetic diffusion and a subgrid model for magnetic dissipation in the form of magnetic reconnection, which can heat the thermal gas.
To enforce the $\nabla \cdot \mathbf{B} = 0$ constraint, we adapted the hyperbolic constrained divergence cleaning of \citet{Tricco2016} (for the implementation into OpenGadget3 see Steinwandel \& Price, in prep.). 

We employed an on-the-fly shock finder \citep[][]{Beck2016a} to compute the necessary shock quantities for CR acceleration. Cosmic rays were represented as populations of protons and electrons and evolved in time using an on-the-fly Fokker-Planck solver, which we discuss in more detail in the next section.

\subsection{Cosmic ray model}

To model CRs, we employed the on-the-fly Fokker-Planck solver \textsc{Crescendo} introduced in \citet{Boess2023} and will only briefly outline the solver and the parameters used here in comparison to the previous work.
\textsc{Crescendo} attaches populations of CR protons and electrons to all resolution elements of the simulation.
These populations are assumed to be isotropic in momentum space and are represented as piece-wise power laws,
\begin{align}
        f(p) = f_i \left(\frac{p}{p_i}\right)^{-q_i}
        \label{eq:piecewise_powerlaw}
,\end{align}
for a number of logarithmically spaced bins.
Protons were considered in the dimensionless momentum range $\hat{p} \equiv \frac{p}{m_p c} \in [0.1, 10^5]$ and electrons in the range $\hat{p} \equiv \frac{p}{m_e c} \in [1, 10^5]$, where $m$ is the mass of the individual particle species.
We discretized the populations with 6 bins (1/dex) for protons and 20 bins (4/dex) for electrons.
This choice of resolution was mainly driven by memory constraints, with the focus on still being able to model accurate high-momentum losses for electrons.
These populations were then evolved in time by solving the diffusion-advection equation in the two-moment approach, following \citet{Miniati2001} \citep[see also][for other recent on-the-fly spectral CR models]{Girichidis2020, Ogrodnik2020, Hopkins2021a}.

We accounted for injection at shocks via DSA, adiabatic changes due to density changes in the surrounding thermal gas, and energy losses of electrons due to synchrotron emission and inverse Compton (IC) scattering off CMB photons.
This reduced the Fokker-Planck equation, which describes the time evolution of our CR populations to
\begin{align}
        \frac{D f(p,\mathbf{x},t)}{Dt} &=  \left( \frac{1}{3} \nabla \cdot \mathbf{u} \right) p \frac{\partial f(p,\mathbf{x},t)}{\partial p} \label{eq:fp-adiabatic} \\
        &+ \frac{1}{p^2} \frac{\partial}{\partial p } \left( p^2 \sum_l b_l f(p,\mathbf{x},t) \right) \label{eq:fp-rad}\\
        &+ j(\mathbf{x}, p, t), \label{eq:fp-sources}
\end{align}
where Eq.~\ref{eq:fp-adiabatic} denotes changes due to adiabatic expansion or compression of surrounding gas, Eq.~\ref{eq:fp-rad} denotes radiative changes due to synchrotron emission and IC scattering off CMB photons with $\sum_l b_l \equiv \left\vert \frac{dp}{dt}\right\vert_\mathrm{syn} + \left\vert \frac{dp}{dt}\right\vert_\mathrm{IC}$, and Eq. \ref{eq:fp-sources} denotes the source term of CRs.
In this work, the source term $j(\mathbf{x}, p, t)$ describes CR acceleration at shocks, modeled following the DSA parametrization by \citet{Ryu2019}.
In addition to that, we used the obliquity-dependent model introduced by \citet{Pais2018} for protons that are primarily accelerated at quasi-parallel shocks, and shifted it by $90^\circ$ for electrons (see Appendix~\ref{app:eta} for more details).
We injected the energy into the full momentum range as a single power law following a linear DSA slope in momentum space,
\begin{align}
    q = \frac{3r}{r - 1}; \quad r = \frac{ ( \gamma + 1 ) \mathcal{M}_s^2}{(\gamma - 1 ) \mathcal{M}_s^2 + 2}
    \label{eq:dsa_slope}
,\end{align}
where $r$ is the shock compression ratio, $\gamma$ is the adiabatic index of the gas, and $\mathcal{M}_s$ is the sonic Mach number.
The DSA parametrization by \citet{Ryu2019} imposes a critical Mach number, $\mathcal{M}_s = 2.25$, below which no efficient CR acceleration can take place.
We employed a fixed electron-to-proton energy injection ratio of $K_\mathrm{ep} = 0.01$.

For this run, we used closed boundary conditions at the lower end of the distribution function, as they provide more numerical stability and mimic low-momentum cooling on adiabatic compression, which is not explicitly included in this simulation.
We solved Eq.~\ref{eq:fp-adiabatic} in the two-moment approach by accounting for CR number and energy changes per bin for both protons and electrons for reasons of numerical stability \citep[see][for a discussion of the benefit for protons]{Girichidis2020}.

Feedback from the CR component to the thermal gas in the form of comoving pressure was calculated by solving the integral
\begin{equation}
    P_{\mathrm{CR},c} = \frac{4 \pi}{3} \: a^{4} \: \int\limits_{p_{\mathrm{min}}}^{p_{\mathrm{cut}}} dp \: p^2 T(p) f(p)
    \label{eq:pressure_integral}
\end{equation}
under the approximation $T(p) \approx p c$.
We started the injection of CRs at $z=4$.
All these effects were computed on the fly for every gas particle active in the timestep.
Constraints on computing memory made it impossible to also account for transport processes, besides advection, which we accounted for naturally, since the CR populations are attached to our Lagrangian SPH particles.
We shall study the effect of transport processes on CRs in galaxy clusters in future zoom-in simulations instead.

\subsection{Fermi-II reacceleration\label{sec:fermi2}}

We implemented an on-the-fly treatment of turbulent reacceleration of CRs and briefly outline the implementation here (for details, see Appendix \ref{app:Dpp}). To follow the turbulent reacceleration of CRs, we adopted the model by \citet{Cassano2005}.
We used the unified-cooling approach, which extends the $b_l(p)$ term in Eq.~\ref{eq:fp-rad} to include the systematic component of the reacceleration,
\begin{equation}
    \sum_l b_l(p) \equiv \left( \frac{\mathrm{d}p}{\mathrm{d}t} \right)_{\mathrm{cool}}^{\mathrm{synch}} + \left( \frac{\mathrm{d}p}{\mathrm{d}t} \right)_{\mathrm{cool}}^{\mathrm{IC}} + \left( \frac{\mathrm{d}p}{\mathrm{d}t} \right)_{\mathrm{acc}}^{\mathrm{sys}} \quad .
\end{equation}
Following \citet{Cassano2005}, the change in momentum due to turbulent reacceleration per timestep can be expressed as
\begin{equation}
        \left( \frac{\mathrm{d}p}{\mathrm{d}t} \right)_{\mathrm{acc}}^{\mathrm{sys}} = -\chi p \approx - 2 \frac{D_{\mathrm{pp}}}{p} = -2 D_0(t) \: p,
        %\label{eq:dpdt_reacc}
\end{equation}
where we used $D_\mathrm{pp} = D_0(t) \: p^2$.

The computation of $D_0(t)$ given in Eq.~\ref{eq:D0} is quite expensive and the way it is coupled to Eq.~\ref{eq:fp-adiabatic} puts a significant timestep constraint on the solver.
We therefore sub-cycled the solver if $D_0(t) \times \Delta t$ surpassed a critical value.
Additionally, we introduced a cap for $D_0(t)$ to avoid overly strong impacts of individual particles on the total runtime of the simulation.

Due to the additional computational cost, we switched this mode off in \crbox $\:$ and ran comparison zoom-in simulations with the mode switched on.
For \comazoom, we switched the mode off to obtain a comparison and see the impact of resolution.
In \comalow $\:$, we ran the turbulent reacceleration on the fly, but capped the value of $D_0$, which limits the need to sub-cycle the solver.
We capped the value at $D_{0,\mathrm{max}} = 10^{-18} \: \mathrm{s}^{-1}$, which lies at the lower end of the values reported in Appendix B in \citet{Donnert2014} but is close to the maximum values found in the Coma filaments of our simulation (we discuss this in Sect. \ref{sec:coma-filament}).
However, this model will need to be revised to include recent results that indicate that preferentially the solenoidal component of turbulence can efficiently reaccelerate CRs \citep[][]{Brunetti2016, Brunetti2020}, or allow for other reacceleration mechanisms \citep[e.g.,][]{Tran2023}.

\subsection{Synchrotron emission}

As in the introductory paper for \textsc{Crescendo}, we computed the synchrotron emissivity, $j_\nu$, in units of [erg s$^{-1}$ Hz$^{-1}$ cm$^{-3}$] as 
\begin{align}
    j_\nu(t) &= \frac{\sqrt{3} e^3}{m_e^2 c^3} \: B(t) \: \sum\limits_{i=0}^{N_\mathrm{bins}} \:\int\limits_0^{\pi/2} d\theta  \sin^2\theta \:  \int\limits_{\hat{p}_\mathrm{i}}^{\hat{p}_\mathrm{i+1}} d\hat{p} \:\: 4\pi \hat{p}^2 f(\hat{p}, t) \: K(x)
    \label{eq:synch_emissivity}
,\end{align}
where  $e$ is the elementary charge of an electron, $m_e$ its mass, $c$ the speed of light, $\hat{p}$ the dimensionless momentum, and $K(x)$ the first synchrotron function,
\begin{equation}
    K(x) = x \int_x^{\infty} dz \ K_{5/3}(z)
    \label{eq:synch_kernel}
,\end{equation}
using the Bessel function, $K_{5/3}$, at a ratio between the observation frequency, $\nu$, and critical frequency, $\nu_c$,
\begin{equation}
    x \equiv \frac{\nu}{\nu_c} = \frac{\nu}{C_\mathrm{crit}  B(t) \sin\theta \: \hat{p}^2}; \quad C_\mathrm{crit} = \frac{3e}{4\pi m_e c} \quad .
\end{equation}
All scalings of the synchrotron emission with frequency and magnetic field strength arise directly from the evolved CR electron distribution function, $f(\hat{p},t)$, in Eq.~\ref{eq:synch_emissivity}, without further assumptions or imposed limits \citep[see Appendix A in][for a more detailed description]{Boess2023a}.

% allsky predictions
\section{Full-sky projections\label{sec:allsky}}
Having a full box simulation of a constrained local Universe puts us in the unique position to study full-sky projections observed from our Milky Way replica. 
To obtain the projections, we mapped the SPH particles of our simulation onto a \textsc{HealPix} sphere \citep[][]{Gorski2005} following the algorithm described in \citet{Dolag2005a}.
All figures in this section were obtained from the output at redshift $z=0$.

\subsection{Cosmic ray electrons}
\begin{figure*}
    \centering
    \includegraphics[width=17cm]{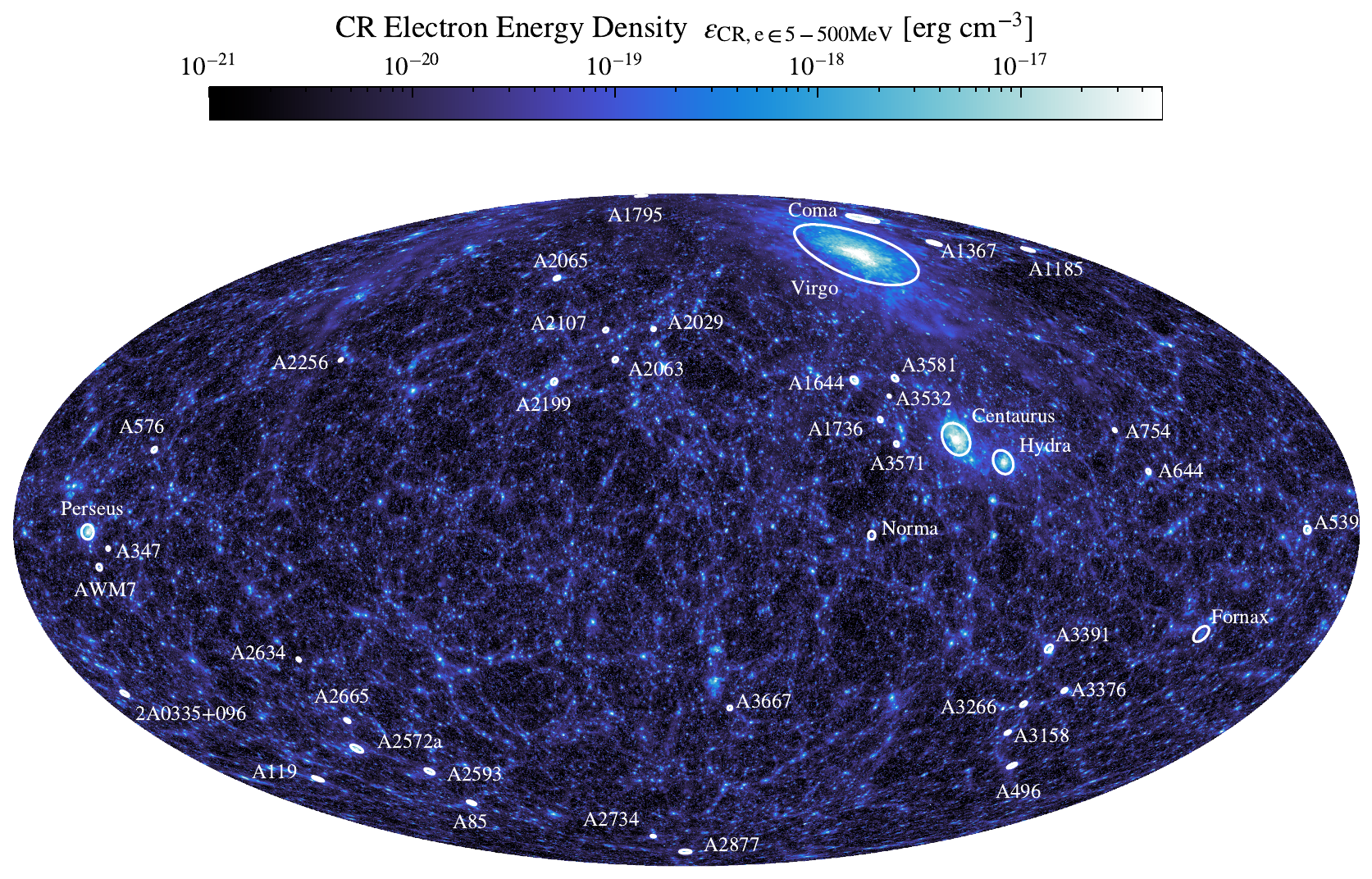}
    
    \includegraphics[width=17cm]{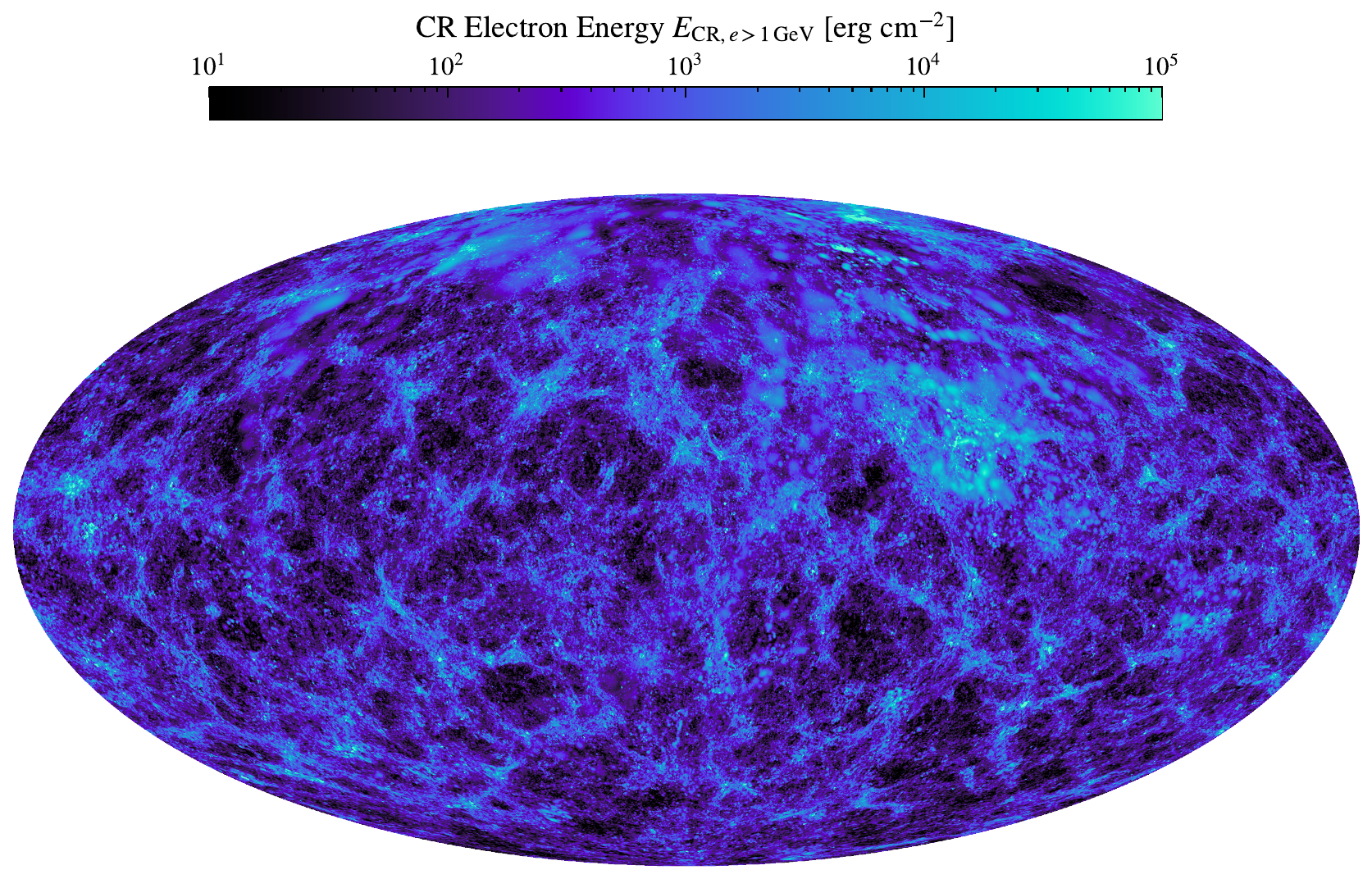}
 %\sidecaption
\caption{Full-sky projections of the CR electron component. Upper panel: Energy density of electrons in the range $E \in [5 - 500]$ MeV as the mean along the line of sight in a radius range $r = 5-300$ Mpc. This traces the injected energy density of the CR electrons with the longest lifetimes. Circles indicate the projected $r_\mathrm{vir}$ of the cross-identified clusters \citep[see][]{Hernandez2024}. Lower panel: CR electrons with energies $E_{\mathrm{CR},e} > 1$ GeV integrated along the line of sight in a radius range $r = 5-300$ Mpc. The short cooling times of these electrons lead to this map tracing recent injection events.}
        \label{fig:CRe_allsky}
\end{figure*}
We show the full-sky projection of our CR electron component in Fig.~\ref{fig:CRe_allsky}.
The labeled clusters are our best matches for the observed counterparts \citep[for details on the matching process, see][]{Hernandez2024}{}{} and the circles indicate their projected virial radii.\\
We show the two opposite ends of the CR electron energy range.
The upper plot shows the mean CR electron energy density along the line of sight in a radius range, $r = 5-300$ Mpc, for CRs in the energy range $E \in [5-500]$ MeV.
This traces the electrons with the longest lifetimes, with a bias toward electrons that have been accelerated and advected into high-density regions.
These electrons provide potential seed populations for reacceleration by subsequent shocks and/or turbulence.

We find a persistent population of CR electrons in galaxy clusters and the cosmic web.
However, this can only be considered to be an upper limit, especially in the highest-density regions, as we do not explicitly treat low-momentum energy losses for CRs.
Hence, this picture should be viewed more as a tracer for shock injection and adiabatic compression and with that a tracer of potential CR electron seed populations.
A distinct feature and advantage of an on-the-fly approach for CR evolution is that we find large-scale halos of electron energy density around clusters and filaments.
This is consistent with recent observations of radio mega-halos \citep{Cuciti2022}, indicating a significant population of CR electrons in the cluster volume.
It has been shown by \citet{Beduzzi2023} that given a turbulent velocity of $v_\mathrm{turb} > \: \sim 150$ km s$^{-1}$ \citep[which is only a factor of 1.5 larger than, for example, the measurements of the atmosphere in Perseus by][]{Hitomi2018} these electrons have almost unlimited lifetimes and with that can remain synchrotron-bright at low frequencies, providing the basis for these radio mega-halos.
We shall discuss the implications of this in future work focusing on radio halos and radio relics in our simulation.

In the lower panel of Fig.~\ref{fig:CRe_allsky}, we show the energy contained in electrons with energies above 1 GeV.
These electrons are potentially synchrotron-bright at frequencies of 144 MHz.
Since the cooling times of these CRs are of the order $\sim 10^2 - 10^3$ Myrs, this predominantly traces recent injection.
We find significant injection at the accretion shocks of most clusters, as well as the accretion shocks around cosmic web filaments.
This shows that accretion shocks around filaments, together with internal shocks, provide a significant, volume-filling source for high energy CRs, available to emit radio emission, given sufficiently strong magnetic fields.

\subsection{Magnetic field\label{sec:Bfield}}
\begin{figure*}
        \centering
    \includegraphics[width=17cm]{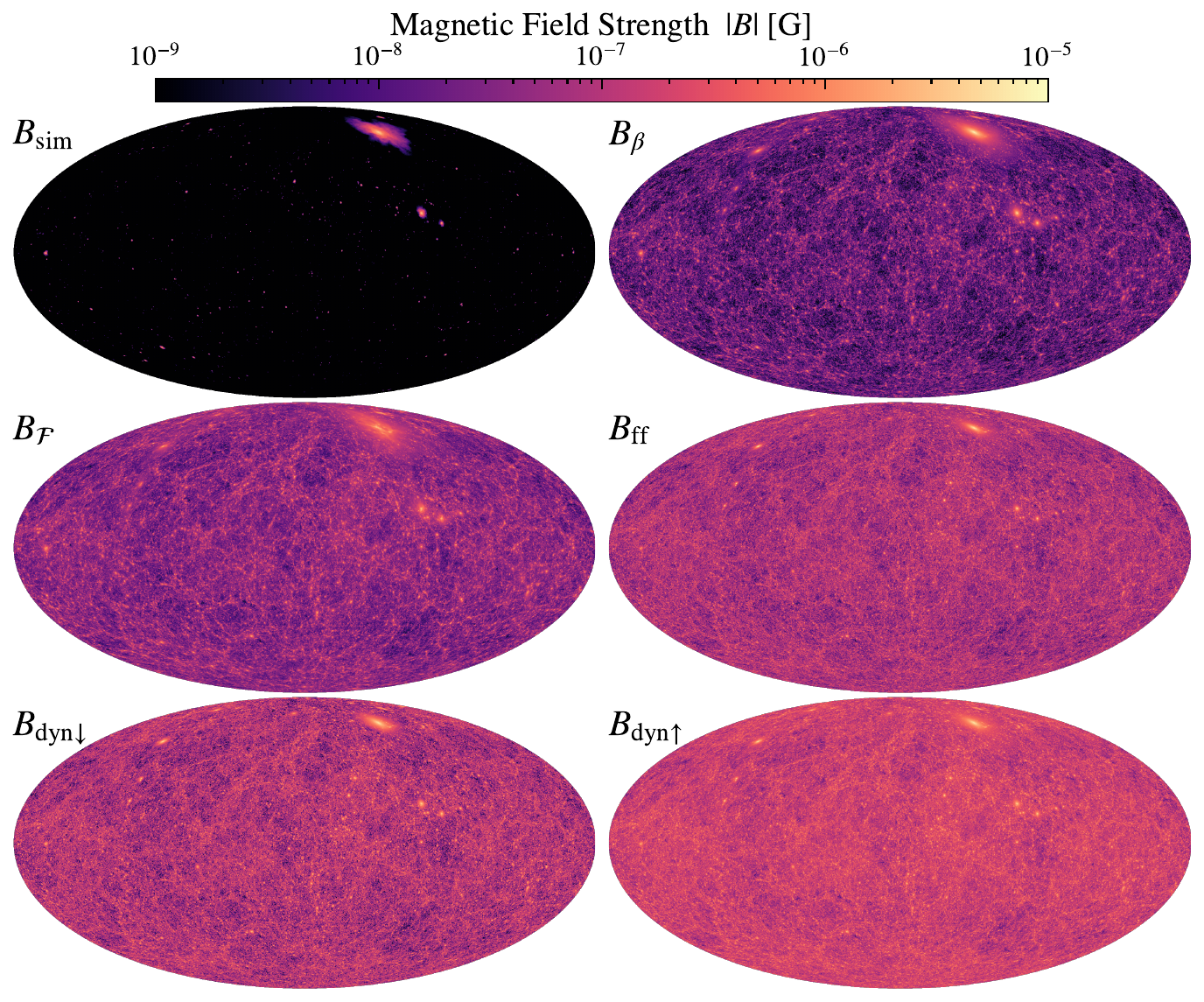}
        \caption{Full-sky projections of the mean magnetic field along the line of sight for the simulated magnetic field and post-processing models. Upper left panel: Simulated magnetic field obtained from the simulation. Upper right panel: Modeled magnetic field, assuming a plasma beta of $\beta \equiv \frac{P_\mathrm{th}}{P_\mathrm{B}} = 50$. Center left panel: Modeled magnetic field based on a fraction $\mathcal{F}$ of the turbulent pressure. Center right panel: Modeled magnetic field based on flux freezing $B \propto \rho^{2/3}$. Lower left panel: Modeled magnetic field, assuming a saturated turbulent dynamo scaling $B \propto \rho^{1/2}$ down to $n_e = 10^{-4}$ cm$^{-3}$ and steeply declining beyond this density to match expectations for void magnetic fields. Lower right panel: Modeled magnetic field, assuming a saturated turbulent dynamo scaling for all density regimes.}
    \label{fig:B_allsky}
\end{figure*}
Since the synchrotron emissivity scales strongly with the magnetic field strength ($ L_\nu \propto B^{\alpha_0 + 1} $, where $\alpha_0$ is the spectral index of the synchrotron spectrum), the observability of synchrotron emission of the cosmic web is tightly linked to the magnetic field strength in filaments.

The magnetic field in our simulation is set up as $\mathbf{B} = (10^{-14}, 0, 0)$ G at the starting redshift $z = 120$.
This field then evolves and is amplified by adiabatic compression and dynamo processes \citep[see e.g.,][]{Stasyszyn2010, Pakmor2014, Marinacci2015, Vazza2014, Vazza2018, Vazza2021, Steinwandel2022_dynamo, 250xMHD}{}{}.
As this is a non-radiative simulation without star formation or cooling, we cannot account for astrophysical sources of a magnetic field that can contribute to volume-filling magnetic fields in low-density regions \citep[][]{Beck2013, Garaldi2021}{}{}.
While we obtain reasonable values in galaxy clusters \citep[of the order $\sim \mu$G, see e.g.,][the latter for a review]{Bonafede2010, Weeren2019}, the magnetic field strength in filaments remains negligible.
We attribute this to a lack of resolution in the filaments to drive an efficient turbulent dynamo to amplify the magnetic field \citep[see][for an extensive study performed with our simulation code]{Steinwandel2022_dynamo}.

To test the implications of different magnetic field strength values in filaments on their potential synchrotron emission, we adopted five magnetic field models, two based on pressure scaling and three based on density scaling.
We show the mean magnetic field along the line of sight in a radius range, $r = 10-300$ Mpc, for the simulation and the scaling models in Fig. \ref{fig:B_allsky}.\\
First, we tested a simple model where the magnetic field pressure is a constant factor of the thermal pressure $P_\mathrm{th}$, such that
\begin{equation}
    B_\beta = \sqrt{\frac{8\pi P_\mathrm{th}}{\beta}} \quad ,
\end{equation}
where we have assumed a constant plasma-$\beta$ of $\beta = 50$.
This is an optimistic model, given that such low values of $\beta$ are typically only found in the central regions of clusters, while realistic values for filaments and accretion shocks are of the order $\beta \sim 10^2 - 10^3$ \citep[see e.g.,][]{Ha2023}.

Second, we calculated the magnetic field as a fraction, $\mathcal{F}$, of the turbulent pressure,
\begin{equation}
    B_\mathcal{F} = \mathcal{F} \sqrt{ 4\pi \rho v_\mathrm{turb}^2 } \quad .
\end{equation}
$\mathcal{F}$ can be estimated from the magnetic Reynolds number following Table 1 in \citet{Schober2015} and takes typical values of $\mathcal{F} \approx 0.01 - 0.1$ in the systems we are interested in.
However, to better match the central values of the magnetic field strength in our Coma analog, we adapted $\mathcal{F} = 1$.
This assumes pressure equilibrium between turbulent and magnetic pressure, as it is expected in a regime where the turbulent dynamo is saturated \citep[see e.g.,][and references therein]{250xMHD}.
Recent results by \citet{Zhou2024} show that shocks are potential regions of efficient magnetic field generation and amplification and can reach this saturation value relatively quickly.
We therefore used the model \bturb~as a maximum assumption of their work.

For the magnetic field models scaling with the density, we first used a simple flux-freezing model where initial magnetic fields are amplified by adiabatic compression in collapsing halos.
Following \citet{Hoeft2008}, we adapted
\begin{equation}
    B_\mathrm{ff} = \frac{B}{0.1 \mu\mathrm{G}} \left( \frac{n_e}{10^{-4} \: \mathrm{cm}^{-3}} \right)^{2/3} \quad ,
\end{equation}
scaling this model up by a factor of four to better match the central magnetic field strength in our Coma analog. We shall discuss this in more detail in Sect.~\ref{sec:coma-filament}.

In ultrahigh-resolution simulations of massive galaxy clusters, \citet{250xMHD} find that the turbulent dynamo is the dominant amplification process, even in low-density regions down to $n_e \sim 10^{-4}$.
From this, we extrapolated two turbulent dynamo models.
The first assumes that the turbulent dynamo amplifies the magnetic field down to densities of $n_e \sim 10^{-4}$ and below that we applied a fit function to intersect the observations by \citet{Carretti2022} with $B \sim 30$ nG at filaments with densities of $n_{e} = 10^{-5} \mathrm{cm}^{-3}$ and extrapolated to $B \sim 10^{-14}$~G at $n_{e} = 10^{-6} \: \mathrm{cm}^{-3}$ as a proxy for void magnetic fields \citep[see e.g.,][for the large differences in the estimation of this value]{Neronov2010, Caprini2015}{}{}.

This gives a model dependent on $n_e$ as
\begin{equation}
    B_{\mathrm{dyn},\downarrow}(\rho) =
    \begin{cases}
        2.5 \times 10^{-6}  \left( \frac{n_e}{10^{-3} \mathrm{cm}^{-3}} \right)^{1/2}& \text{[G],  if $n_e > 10^{-4}$ cm$^{-3}$}\\
        \sum\limits_{i=1}^5 p_i \log_{10}(n_e)^{i-1}  & \text{[$ \log_{10}$ G],  if $n_e < 10^{-4}$ cm$^{-3}$}
    \end{cases}
,\end{equation}
where $p \in [-16.38, -16.0, -8.07, -1.71, -0.13]$.\\
%This model can be considered as a lower limit.\\
%
We also considered an upper limit, as a pure saturated dynamo scaling with density as
\begin{equation}
        B_{\mathrm{dyn}, \uparrow}(\rho) = 2.5 \times 10^{-6}  \left( \frac{n_e}{10^{-3} \mathrm{cm}^{-3}} \right)^{1/2} \text{[G]} ,
\end{equation}
where we fit the results from \citet{250xMHD} and extended the scaling to lower densities with $B \sim 250$ nG at $n_{e,0} = 10^{-5} \mathrm{cm}^{-3}$.
Identical to the previous scaling, this matches magnetic field strengths in cluster centers very well, but overestimates the magnetic field strength in filaments by roughly one order of magnitude, compared to the \citet{Carretti2022}.
This is, however, consistent with previous works by \citet{Vernstrom2017}, \citet{Brown2017}, \citet{OSullivan2019}, \citet{Locatelli2021}, who allow for up to 250 nG.
To illustrate the scaling of our models as a function of electron density, we refer to Fig.~\ref{fig:Bmodels}.
\subsection{Synchrotron emission}
\begin{figure*}
        \centering
        \includegraphics[width=17cm]{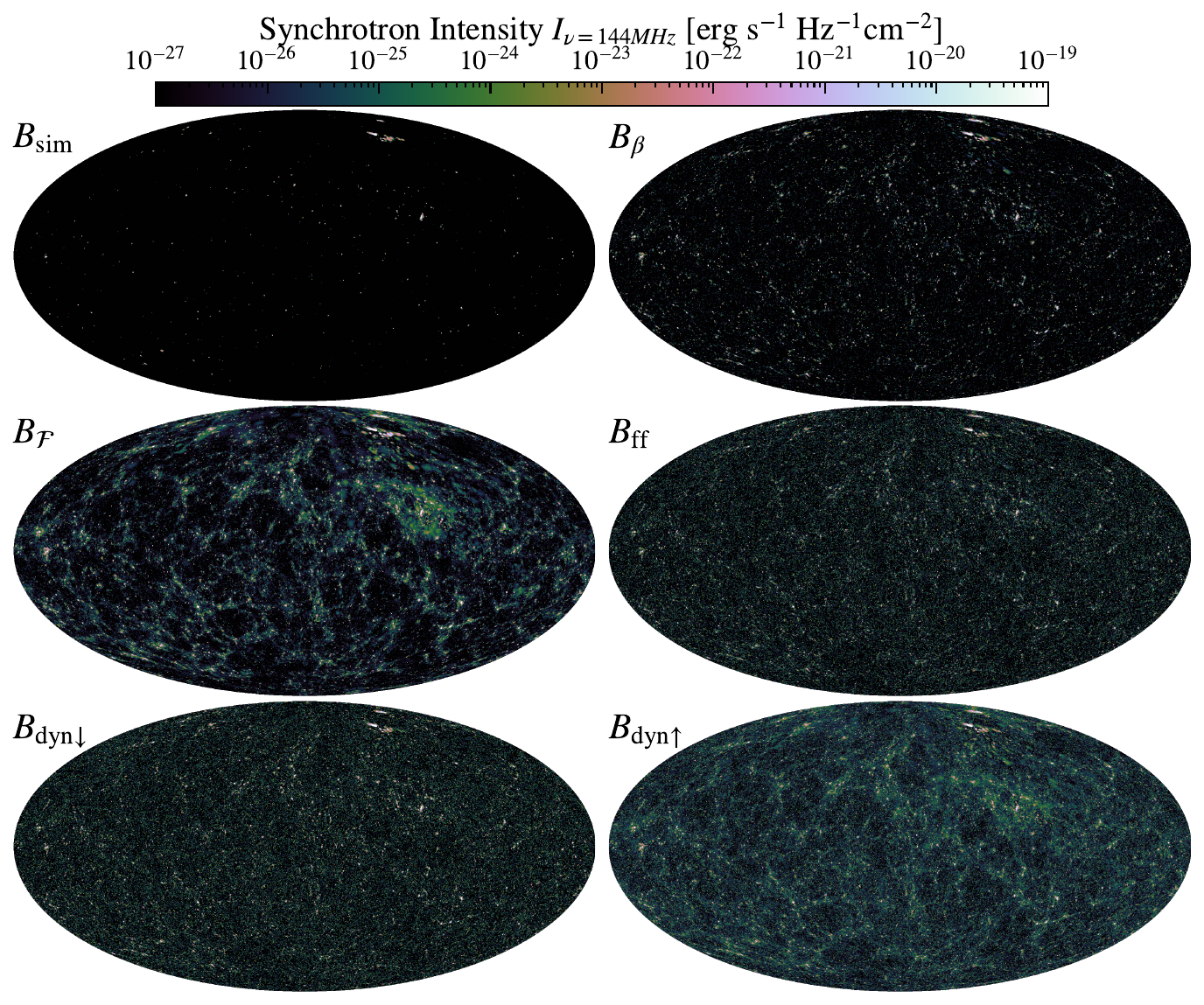}
        \caption{(Intrinsic) synchrotron intensity at 144 MHz integrated along the line of sight between $r = 10 - 300$ Mpc. The different panels show the emission calculated based on the magnetic field configurations shown in Fig. \ref{fig:B_allsky}.}
    \label{fig:synch_allsky}
\end{figure*}
By folding the lower panel of Fig.~\ref{fig:CRe_allsky} with the different panels of Fig.~\ref{fig:B_allsky} according to Eq.~\ref{eq:synch_emissivity}, we can obtain the synchrotron emissivity.
From there, we arrive at the intrinsic synchrotron surface brightness by integrating the emissivity of the SPH particles along the line of sight.
We show the result of this for the previously introduced magnetic field models in Fig.~\ref{fig:synch_allsky}.

In the case of the simulated magnetic field, we only obtain synchrotron emission from galaxy clusters.
The emission in cluster centers is of the order $\sim$ten lower compared to observations, which is mainly driven by the flatter radio spectra.
Our simulated spectra in the box stem purely from direct injection and advection, so they lack additional steepening from on-the-fly turbulent reacceleration.
We discuss this in more detail for the filaments of Coma in Sect.~\ref{sec:coma-filament} and for the radio emission in Coma, Virgo, and Perseus in Dolag et al. (in prep).

The lack of emission in the filaments stems from the steeper dropoff of the magnetic field as a function of density.
As was discussed before, the simulated magnetic field strength in the filaments is well below the nano-Gauss level and with that would require substantial CR injection to be synchrotron-bright.

Assuming a magnetic field following a constant plasma-$\beta$ (\bbeta) yields more interesting results.
We still primarily get synchrotron emission from the clusters; however, we also see more emission by smaller clusters that do not have a strong enough simulated magnetic field to be synchrotron-bright.
In addition, we also see more relics far from the cluster center, such as in our Perseus replica to the very left of the map.
This stems from the magnetic field scaling with the thermal pressure and with that automatically tracing regions of over-pressure; namely, shocks.
We can observe some filamentary structure in the synchrotron emission; however, this is mainly driven by the halos inside the filaments, rather than diffuse emission from filaments.

The magnetic field scaling with turbulent velocity (\bturb) produces morphologically similar features as the constant plasma-$\beta$ case.
This naturally originates from shocks inducing a substantial amount of turbulent energy \citep[e.g.,][]{Ryu2003, Vazza2011}, so this model again amplifies magnetic field strength and, with that, synchrotron emission at shocks.
This provides even more synchrotron emission from accretion shocks around clusters than in the \bbeta-model and the first indication of synchrotron emission from accretion shocks around cosmic web filaments.

The density-dependent magnetic field models \bff~and \bdynl both show similar behavior.
These models lead to preferential emission in the center of halos with a steep decline toward lower densities.
This leads to filamentary emission driven by halos within the filaments, but no diffuse emission from the gas in filaments or accretion shocks.

\bdynh is the only model that produces considerable diffuse synchrotron emission from filaments.
We find consistent diffuse emission within the filaments and in some regions in the central and top right parts of the map, the magnetic field is even strong enough to illuminate part of the accretion shocks onto the clusters.
The emission by filaments and accretion shocks have intensities more than five orders of magnitude lower than the intensities in cluster centers, making prospects of direct detection very difficult, as was expected.

Further, from a numerical standpoint, it is difficult to capture the spurious shocks in the low-density environment around filaments which can lead to an underestimation of the synchrotron emission in these maps.
This mainly stems from the numerical resolution in these low-density regions.
We discuss the impact of resolution on our findings in Sect.~\ref{sec:coma-filament}.

% filaments
\section{The Coma filaments}
\label{sec:coma-filament}
\begin{figure*}
    \centering
    \includegraphics[width=17cm]{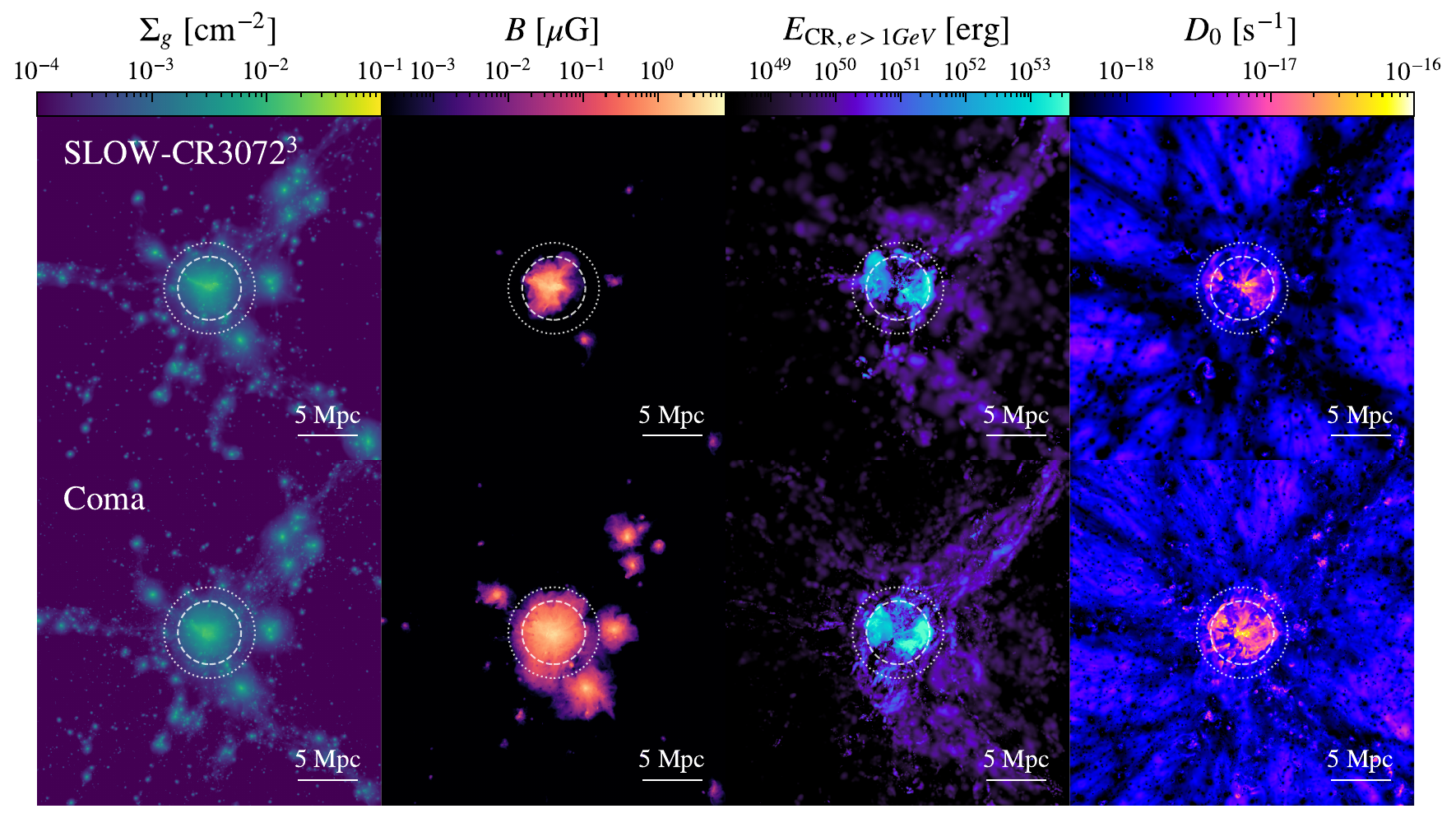}
    \caption{Cutout around our Coma replica with a side-length of $d = 20 h^{-1}c$Mpc. The circles represent $r_{200}$ and $r_{500}$. 
    From left to right, we show surface density, absolute value of the simulated magnetic field, CR electron energy of electrons with energies above 1 GeV, and turbulent reacceleration coefficient. The upper panels show the cluster in the full cosmological box; the lower panels show it in the zoomed-in re-simulation.
    }
    \label{fig:coma_resolution}
\end{figure*}
To study the potential synchrotron emission from cosmic web filaments in more detail, we shall now focus on the Coma cluster replica found in our simulation and its surroundings.
The cosmic web filaments around our Coma replica have recently been studied in \citet{Malavasi2023}, where they find good agreement with indications from observations, which provides a reasonable basis for our analysis.
\subsection{Impact of resolution}
To test the impact of resolution on our results, we added to our analysis by presenting results from zoom-in simulations of the Coma cluster (\comazoom \& \comalow) at eight times the resolution of the cosmological box (\crbox).
We show the comparison between \crbox~and \comazoom~in the top and bottom panels of Fig.~\ref{fig:coma_resolution}.
From left to right, we show surface density, the absolute strength of the simulated magnetic field, the CR electron energy of the electrons with energies above 1 GeV, and the turbulent reacceleration coefficient given in Eq.~\ref{eq:D0}.

We find excellent morphological agreement between the different setups, as can be seen in the density plot in the left panels, providing us with a good basis for comparison.
In Coma itself, we can also see reasonable agreement in the magnetic field strength (second panels), injected CR energy (third panels), and reacceleration coefficient (last panels).

However, as we move away from Coma, we can observe the impact of resolution.
In the magnetic field strength, this is evident as the magnetic field retains higher values further from the cluster center in the case of the zoom-in simulation. 
As was discussed previously, we attribute this to the lack of resolution to drive an efficient amplification via the turbulent dynamo in \crbox, as is shown in \citet{Steinwandel2022_dynamo}.

Another impact of resolution can be seen in the CR electron panels, where the increased resolution in the zoom-in simulation leads to a more accurate shock capturing and with that improved CR injection.
This can be seen both in the strong shock of the ongoing merger, as well as in the surrounding medium.
In the zoom-in simulation, even the equatorial shock from the current merger is captured, with the cluster in the full box only showing a hint of this shock in direct comparison.\\
Both simulations show an envelope of a previous shock moving away from the center and colliding with the accretion shock \citep[see e.g.,][for a high-resolution study of this process]{Zhang2020a, Zhang2020}.
We find a prominent acceleration of CRs at a filament to the top right of the cluster in both simulations.
This is, however, better refined in the zoom-in simulation, where a clear filament can be seen in the CR component, while the box simulation shows more CR injection on one side of the filament than the other.

\subsection{Fermi-II reacceleration}

The turbulent reacceleration coefficient maps in the rightmost panels of Fig. \ref{fig:coma_resolution} again agree reasonably in morphology and absolute value.
We find strong potential for reacceleration in the post-shock regions of the merger and the center of the cluster.
Here values reach up to $D_0 \sim 10^{-16} \: \mathrm{s}^{-1}$, in agreement with the maximum reported in \citet{Donnert2014}.
This shows how conservative a cap at $D_0 = 10^{-18} \: \mathrm{s}^{-1}$ is in the case of a massive galaxy cluster.
However the filaments only reach values of $D_0 \sim  3 \times 10^{-18} \:\mathrm{s}^{-1}$.
Since we focus on filaments in this work we accept this limitation for the benefit of substantially lower run-time of the simulation.
\subsection{Magnetic field}
\begin{figure*}[t]
    \centering
    \includegraphics[width=\textwidth]{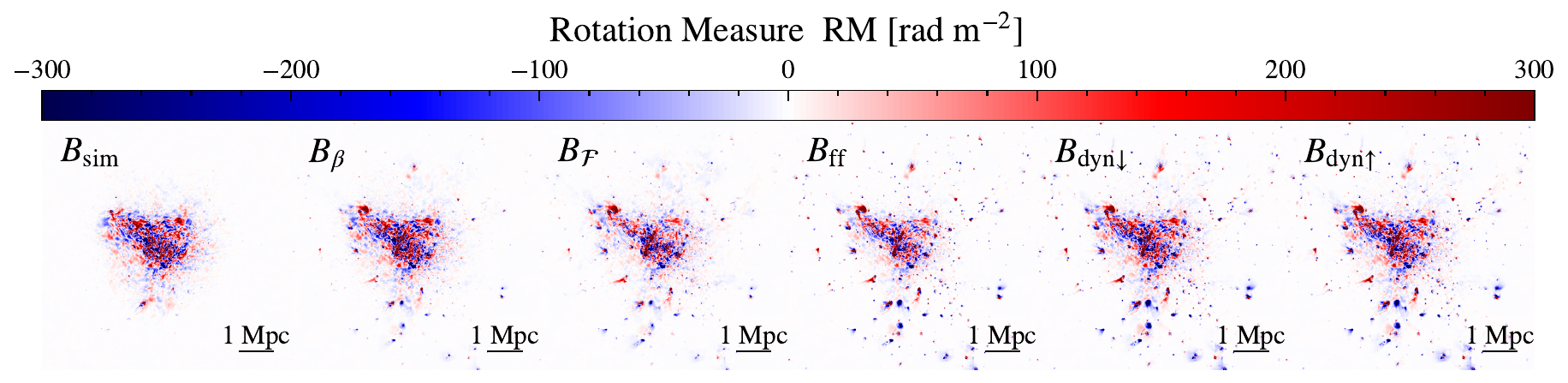}

    \includegraphics[width=\textwidth]{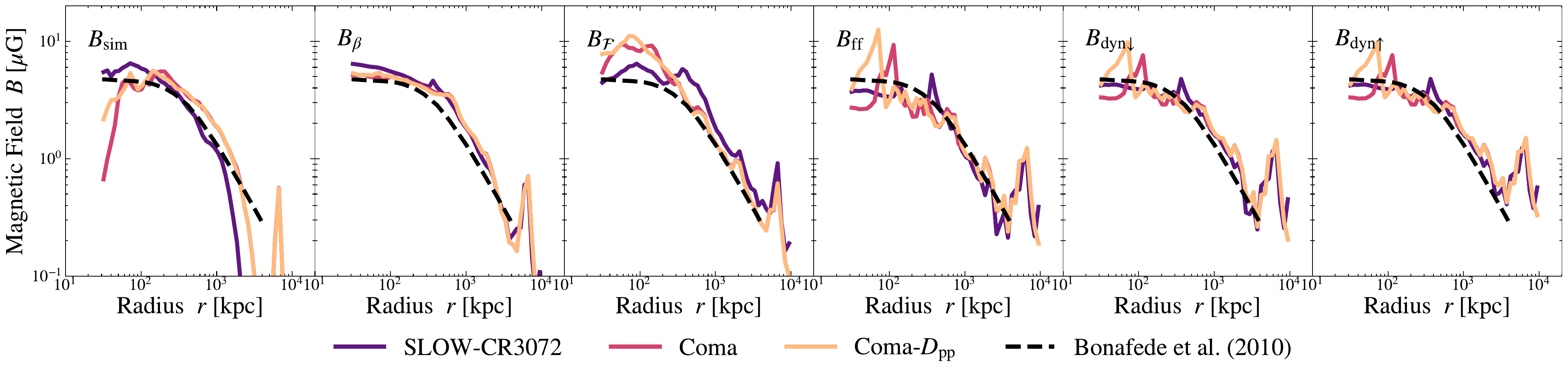}
    \caption{Magnetic field models applied to the \comazoom~simulation. Top: Rotation measure (RM) mock observations of our Coma replica assuming different magnetic field models. Bottom: Radial profiles of the magnetic field. We show the profiles of all discussed simulations for the various magnetic field models in the solid colored lines. The dashed black lines show the observed profile by \citet{Bonafede2010}.}
    \label{fig:Bfield_radial}
\end{figure*}
In Fig.~\ref{fig:Bfield_radial}, we show RM mock-observations and radial profiles of the magnetic field strength for our Coma replica.
The top panels show the RM mock observations where we rescaled the simulated magnetic field to match our magnetic field models in Sect.\ref{sec:Bfield}.
the bottom panels show the radial profiles of the total magnetic field strength.
Different colors correspond to the three simulations, and dashed black lines show the magnetic profile derived from the observations of Coma presented in \citet{Bonafede2010}.

For the simulated magnetic fields, all simulations are in reasonable agreement with the observed profile up to a radius of $\sim 1$ Mpc.
Beyond this radius, the magnetic field profile drops off too steeply, which we attribute to the lack of resolution in the cluster outskirts to resolve the turbulent dynamo amplification, as was mentioned above.

The magnetic field model based on a constant plasma-$\beta$ (\bbeta) similarly agrees well with observations within the inner Mpc, after which the magnetic field is overestimated due to the unrealistically large energy density ratio of the magnetic field to the thermal pressure in the cluster periphery.

Our model based on the fixed ratio between turbulent and magnetic pressure (\bturb) overestimates the central magnetic field strength by roughly 50 percent, but agrees very well with the radial decline in the cluster outskirts.
As mentioned above, this upscaled model is the maximum assumption with pressure equipartition between turbulent kinetic and magnetic pressure and should be considered more as an upper limit.

We applied a similar up-scaling to the \bff~model, which yields good matches between modeled and observed magnetic field strength.

Both turbulent dynamo scaling models (\bdynh~\& \bdynl) agree well with the central value of the magnetic field strength, but differ in the outskirts of the cluster.
\bdynl~transitions to a steeper density scaling at roughly the virial radius and with that a steeper radial decline.

\bdynh~retains its $\sqrt{\rho}$-scaling and with that diverges from the observed radial decline, leading to an overestimation of the magnetic field in regions with $n_e < 10^{-4} \: \mathrm{cm}^{-3}$.
\subsection{Synchrotron flux}
\begin{figure*}
    \centering
    \includegraphics[width=17cm]{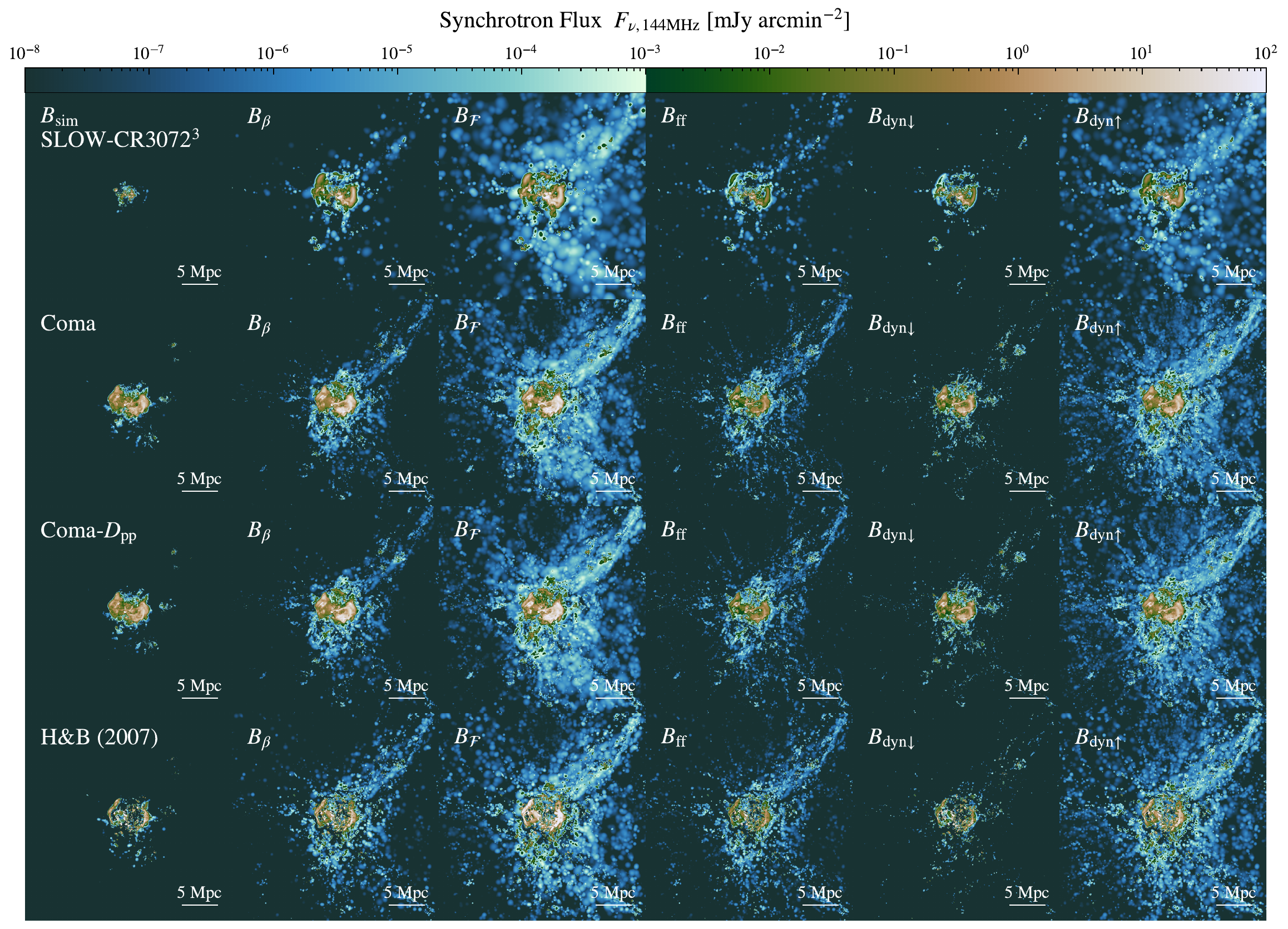}
    \caption{Same cutout as in Fig. \ref{fig:coma_resolution}, but showing the synchrotron surface brightness at 144 MHz, assuming a LOFAR beam size of $\theta = 60" \times 60"$ as in \citet{Bonafede2022}. The color bar is split at $1 \: \mu$Jy beam$^{-1}$, which is a factor two above the sensitivity limit in the stacking approach by \citet{Hoang2023}.
    From top to bottom, we show the different simulation runs, and from left to right we vary the magnetic field models.
    The lowest row shows the result of painting on synchrotron emission with the \citet{Hoeft2007} model in a post-processing approach.
    For this, we use the \comazoom~simulation due to the better performance of the shock finder at this resolution, compared to \crbox.}
    \label{fig:coma_synch}
\end{figure*}
In Fig.~\ref{fig:coma_synch}, we show the expected synchrotron flux of our simulated Coma cluster and its environment, assuming a LOFAR beam size of $\theta = 60" \times 60"$ \citep[as in Fig. 2 of][]{Bonafede2022}.
We convert the intrinsic radio power $P_\nu$ to an observed flux by placing our Coma replica at a redshift of $z=0.0231$ and converting it via $S_\nu = \frac{P_\nu}{4\pi D_L}$ where we use our intrinsic \citet{Planck2014} cosmology for the computation of the luminosity distance $D_L$.
We find that our simulated relic is 1-2 orders of magnitude less luminous than the observed one.
We attribute this first to the possible connection of the relic to a narrow-angle tail galax in the real Coma cluster \citep[e.g.,][]{Bonafede2022, Churazov2022}, which cannot be reproduced here since we do not include galaxy formation and AGN jet physics in this simulation.
A second reason is the known discrepancy between current DSA models and the up to 10 percent efficiency required to reproduce radio relic brightness \citep[see discussion in][]{Botteon2020}.
We discuss the impact of CR injection parameter choices in more detail in Sect.~\ref{sec:discussion}.
For this reason, we rescale the simulated synchrotron flux by a factor of 100, to roughly match the observed brightness of the radio relic in the zoom-in simulations using the simulated magnetic field.
As the acceleration efficiency of CR electrons is still under debate in the literature, this instead allows us to discuss the synchrotron flux in the filaments relative to the flux in the relics.

The top panels show the Coma replica in the cosmological box (\crbox), the second row shows the cluster in the zoom-in simulation without turbulent reacceleration (\comazoom) and in the third row we include turbulent reacceleration where we cap the value of $D_0$ at $D_{0,\mathrm{max}} = 10^{-18} \:\mathrm{s}^{-1}$ (\comalow).
From left to right we show the flux from the simulated magnetic field and the different magnetic field models introduced in Sect.~\ref{sec:Bfield}.

We shall focus on the results of our on-the-fly treatment of CRs here and discuss the comparison to the \citet{Hoeft2007} model shown in the lowest panels of Fig.~\ref{fig:coma_synch} in Sect.~\ref{sec:HB07}.\\
The relic replica in our simulation is only really visible in the zoom-in simulations with the simulated magnetic field. 
As is shown in Fig. \ref{fig:Bfield_radial}, we find that the simulated magnetic field drops off too steeply as a function of radius and with that is not strong enough to illuminate the relic in the lower resolution simulation, even though the CR injection is virtually identical, as is seen in Fig. \ref{fig:coma_resolution}.
This drop in magnetic field is even more pronounced in the filaments and with that does not provide diffuse synchrotron emission of any considerable value.

Applying the magnetic field model based on a constant plasma-$\beta$ (\bbeta) yields more synchrotron emission in the surrounding medium of the cluster for all simulations.
In the case of \crbox~this gives a glimpse at the synchrotron flux of the accretion shock onto the filament in the top right corner of the image.
However, it is very dim due to our shock finder being resolution-limited.
With the zoom-in simulations, this accretion shock is more pronounced and brighter by roughly one order of magnitude, due to the more accurate shock capturing and with that more consistent acceleration.
The inclusion of turbulent reacceleration in \comalow~does not have any considerable impact on the synchrotron flux, due to the small value of $D_\mathrm{pp}$ in the filament region.
Remnants of a shock from a previous merger, perpendicular to the ongoing merger, can be seen as diffuse emission around the cluster in the zoom-in simulations visible as very light blue regions to the north and south of the cluster.

As was the case with Fig. \ref{fig:synch_allsky} the model \bturb produces morphologically very similar results to \bbeta.
Similar to \bbeta~the model \bturb~gives increased magnetic field strength at shocks, which boosts the synchrotron emission at ongoing merger- and accretion shocks, compared to the simulated magnetic field.
With this model even the accretion shock is visible, enveloping the cluster at a radius of roughly 5Mpc.
Studying the time evolution of the cluster we find that the accretion shock has been pushed out by the previous merger activity, consistent with the findings by \citet{Zhang2020a, Zhang2020}.
Further, we find a secondary shock at a radius of roughly 10Mpc, visible to the right of the cluster in the \comazoom~and \comalow~simulations akin to a runaway shock described by the same authors.

With its steeper density-scaling \bff~does not significantly boost magnetic field strength in the density regimes of accretion shocks and filaments, hence the simulated emission is generally very low.
We can again see the impact of resolution between \crbox~and \comazoom~where the improved shockfinder performance increases the simulated synchrotron flux.

For \bdynl~this is even more pronounced, leading to no considerable synchrotron emission in any of the simulations.
Given that this model is the only one tuned to match the most recent observations by \citet{Carretti2022} for magnetic field strength in filaments, it makes the detection of diffuse emission very unlikely with current instruments.

\bdynh~generally provides the strongest magnetic field in the filaments, which directly translates to the most synchrotron flux in the case of this model.
In \crbox~we again see some of the accretion shock onto the top right filament, while in the zoom-in simulations, this is the magnetic field model that provides some tangible diffuse emission from the filaments connecting to Coma.
For \comazoom~, this flux reaches $F_\nu \sim 0.01-1$ $\mu$Jy beam$^{-1}$ and with that lies below current observational capabilities by 0.5-1.5 orders of magnitude compared to the noise level achieved in \citet{Hoang2023}.
Including the modeling of turbulent reacceleration of CRs, with a cap at $D_0 = 10^{-18}\:\mathrm{s}^{-1}$ does not change the synchrotron emission significantly in any of the magnetic field models.
\subsection{Resolved synchrotron spectrum}
\begin{figure*}[t]
    \centering
    \includegraphics[width=17cm]{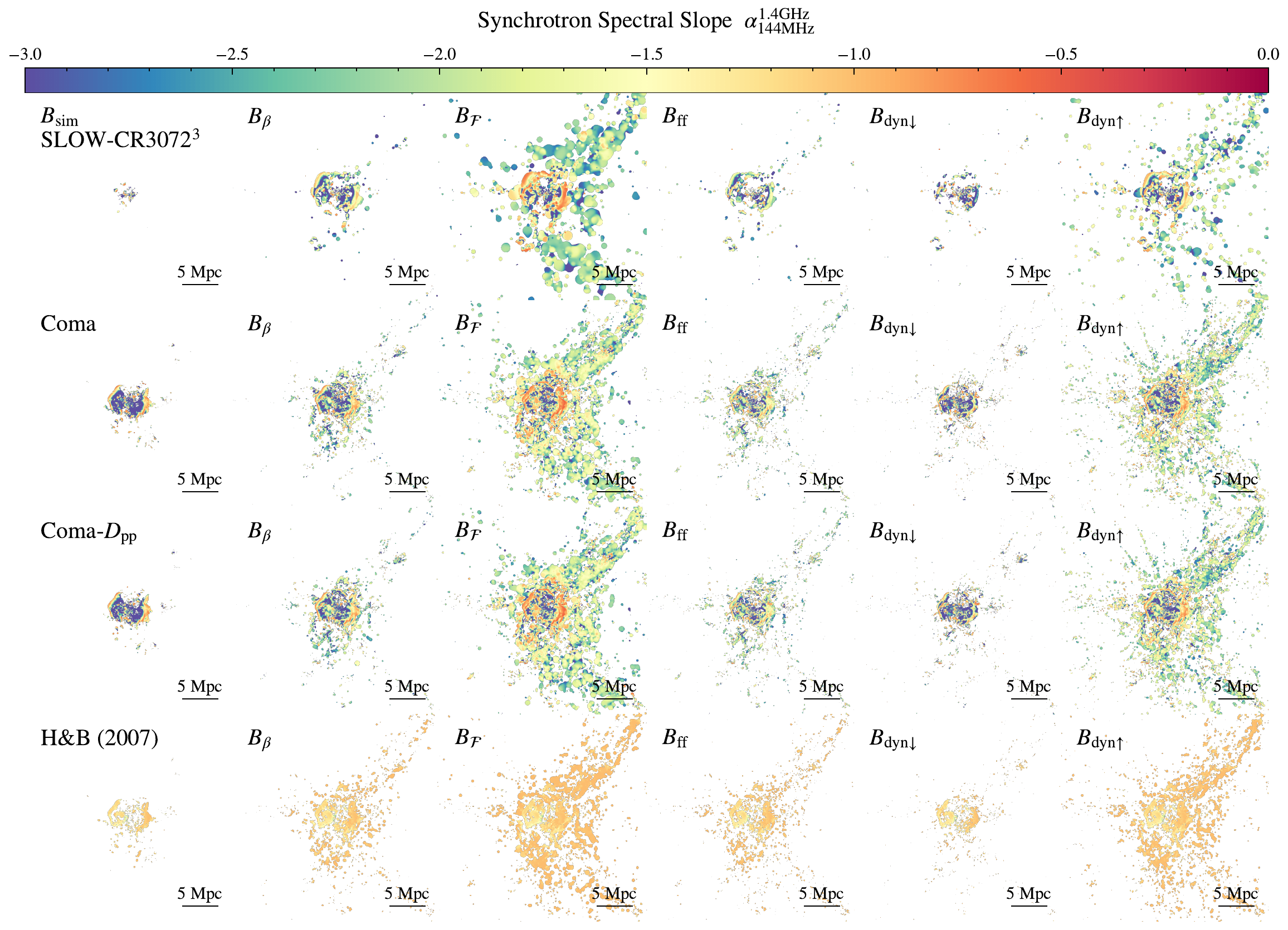}

    \includegraphics[width=17cm]{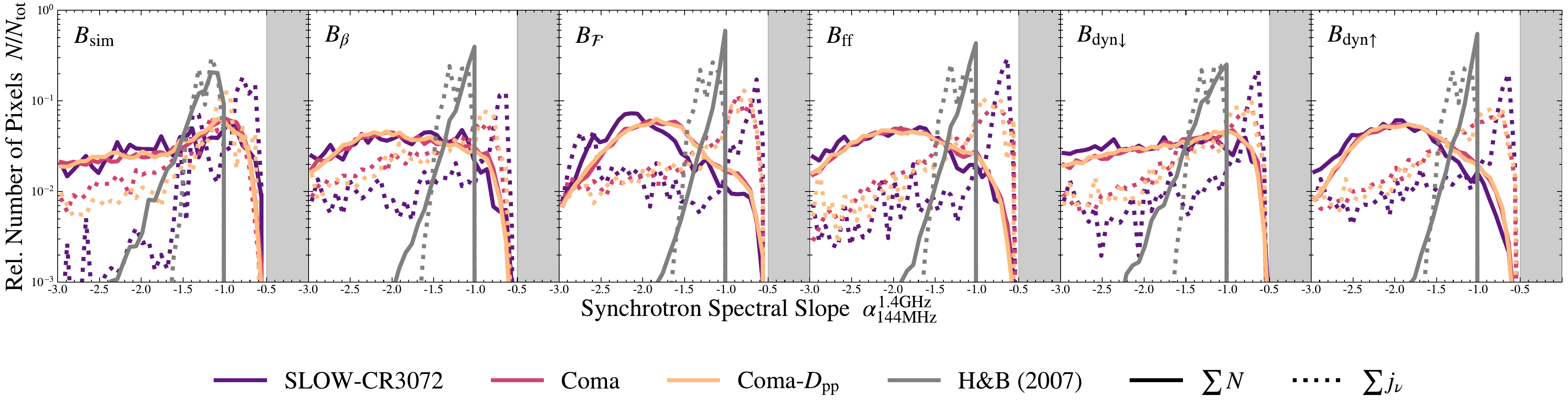}
    \caption{Resolved synchrotron spectrum maps accompanying Fig. \ref{fig:coma_synch}. We constructed the synchrotron slope between the images of Fig.~\ref{fig:coma_synch} and images with the same resolution at 1.4 GHz by fitting a single power law between the respective pixel values. We impose a surface brightness cutoff for the slope construction, similar to the one in Fig.~\ref{fig:coma_synch} and also cut pixels with slopes $\alpha < -3$.
    The lowest row shows histograms of the pixels in the spectral slope maps. Solid lines are simple histograms, dotted lines are synchrotron flux-weighted histograms. The grayed-out area indicates spectral slopes that are too flat for the standard DSA picture.}
    \label{fig:coma_synch_slope}
\end{figure*}
The simulation of the CR electron spectra allows us to self-consistently model synchrotron spectra from the injected and aged electron populations.
In Fig.~\ref{fig:coma_synch_slope}, we show the resolved spectral index maps between 144 MHz and 1.4 GHz.
We constructed these spectra by fitting a single power law to the surface brightness per pixel of the images in Fig.~\ref{fig:coma_synch} and images with the same resolution obtained at a frequency of 1.4 GHz as
\begin{equation}
    \alpha_\mathrm{144 MHz}^\mathrm{1.4 GHz}(\mathrm{pix}) = \frac{\log_{10}(I_\mathrm{144 MHz}(\mathrm{pix})) - \log_{10}(I_\mathrm{1.4 GHz}(\mathrm{pix}))}{\log_{10}(144 \: \mathrm{MHz}) - \log_{10}(1.4 \: \mathrm{GHz})}.
\end{equation}
We impose a surface brightness cutoff similar to the one in Fig.~\ref{fig:coma_synch}, beyond which we do not consider pixels for the slope calculation.
This helps avoid polluting the image with extremely low surface brightness emission that nonetheless can show a flat spectrum due to recent injection at high Mach number accretion shocks.
In addition, we exclude spectral indices smaller than $\alpha < -3$, also purely for visualization reasons.

The spectral index maps show two main behaviors. First, we find the typical spectral steepening in the wake of the ongoing merger shocks. This stems from the synchrotron and IC cooling the CR electrons experience after the shock passage. Second, in the most optimistic magnetic field models which show enough diffuse synchrotron emission from the filaments to model a synchrotron slope, we find a spectral index for diffuse emission in filaments of $\alpha \sim$ -2.0 -- -2.5.
This is steeper than measurements from recent observations by \citet{Vernstrom2023}.
The main reason for this is a choice in numerical design: the injection of CRs into our model happens in the middle of the numerical timestep, in order to subtract the injected CR energy from the entropy change in the hydro solver. 
This is done to conserve the total energy and for numerical stability of the computation \citep[see the discussion in][]{Boess2023}.
All other CR operations, adiabatic changes, turbulent reacceleration, and most relevant here, radiative losses are computed at the end of the timestep for the whole timestep.
Snapshots are then written at the beginning of the subsequent timestep.
This means that a spectrum that has cooled over at least one time step is written in a snapshot.
Due to the adaptive nature of the time-stepping in \textsc{OpenGadget3}, particles in low-density regions such as filaments are always evolved with larger timesteps than those in high-density regions, such as clusters and shocks in clusters.
Due to the fast cooling by IC scattering off the CMB alone, spectra of CRs injected at accretions shocks in filaments will be artificially more cooled than those of CRs accelerated at merger shocks.

The second reason is the wide frequency range over which we consider the resolved synchrotron spectrum in this figure.
Going to lower frequencies, below 200MHz, where the synchrotron spectrum is less affected by cooling processes and does therefore probe the injection-spectrum, we find an excellent agreement with \citet{Vernstrom2023}, as we discuss in Sect.~\ref{sec:filament_spectrum}.
With our imposed limits for slope construction, however, we only find sufficient diffuse emission in the filament to perform this slope construction for the magnetic field models \bbeta, \bturb~and \bdynh.

The lowest panels show the histograms of the synchrotron slope per pixel in the above maps.
Solid lines show the simple histograms, while dotted lines show synchrotron flux-weighted histograms.
Line colors correspond to the different simulations, as is indicated below the figure.
The grayed-out area indicates the range of synchrotron slopes that are too flat to be consistent with DSA.
We note that even though some lines seem to be crossing into this area because of their width we find no pixels in the actual maps that show a synchrotron spectrum flatter than $\alpha = -0.5$, the theoretical limit of DSA.

These histograms generally show a broad distribution of spectral slopes, with the peaks at $\alpha > -1$ being driven by the ongoing injection in the relics, as can be seen from the synchrotron flux-weighted histograms.
For the magnetic field models with strong fields in the filaments and simulations including turbulent reacceleration, we can see a large number of pixels with steeper synchrotron spectra, originating from the diffuse emission around the cluster and in filaments, as was discussed above.

\subsection{Filament spectrum\label{sec:filament_spectrum}}
\begin{figure*}
    \centering
    \includegraphics[width=17cm]{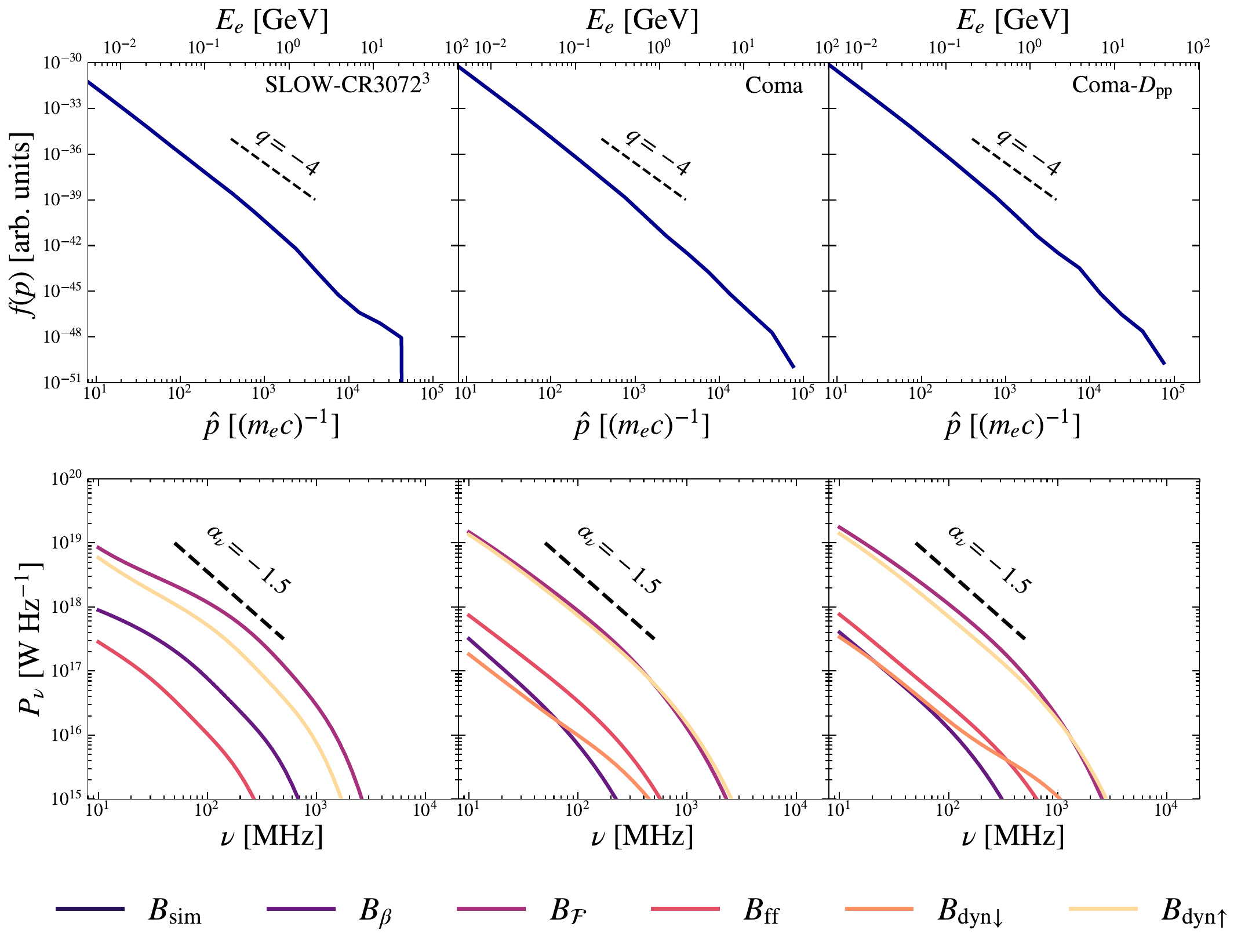}
    \caption{Momentum and synchrotron spectra of CR electron populations in the filament segment indicated in Fig.~\ref{fig:spectra_selection}. Top panels: Sum of all momentum spectra for particles contained in the selected region. Bottom panels: Synchrotron spectra of the selected regions under the assumption of varying magnetic field models, as is indicated by the colors. In all cases, the simulated magnetic field strength is too low to produce synchrotron emission visible in the plotted range.}
    \label{fig:synch_spectrum}
\end{figure*}
To give a more quantitative impression of the synchrotron spectra arising from the CR electron spectra present in the filaments we plot these spectra in Fig.~\ref{fig:synch_spectrum}.
The position of the selected filament region is sketched in Fig.~\ref{fig:spectra_selection}.

The upper panels show the sum of the momentum spectra within the selected region.
These spectra are naturally injection-dominated by high Mach-number accretion shocks around the filament and therefore lie close to the maximum possible DSA slope of $q=-4$.
We can see the impact of resolution and with that, shockfinder performance discussed above in the comparison between the left-most panel and the panel to the right of it.
The larger normalization of the spectrum is driven by a larger number of SPH particles containing a CR population, leading to a larger sum of the CR number and energy per bin.
In the upper right panel an indication of the impact of turbulent reacceleration can be found in the larger normalization at the low-momentum end of the distribution function.
This difference is however very small and can also be explained by a slight difference in the evolution of the cluster.

The synchrotron spectra (shown in the lower panels) on the other hand are dominated by the magnetic field strength.
In all cases, the simulated magnetic field strength is too low to result in synchrotron power large enough to appear in our plotted range.
For our assumed models the magnetic field will always be stronger in the central regions of the filament, where the electrons will settle after some time.
During this time the electrons will cool primarily due to IC scattering off CMB photons.
Hence the high-frequency end of the synchrotron spectra is steeper than one might expect from the sum of the electron spectra.
In the low-frequency range $\nu < 200$ Mhz, the synchrotron spectra are in excellent agreement with the slope $\alpha \approx$ -1 -- -1.5, as is indicated by the dashed black reference lines.
In this range, the CR spectra are not very affected by cooling, so it gives a closer representation of the injection spectra, as was discussed above.

For the \crbox~simulation we find the steepest spectra with the lowest synchrotron power, which we attribute to a lack of resolution to capture spurious shocks around the filament.
This leads to this spectrum being driven by older CR populations which have advected downstream of the acceleration zone and have undergone considerable cooling. In the case of \comazoom~we see similar behavior; however, the total emission is larger, due to the increased numerical performance of our shock finder and with that more injection of CRs.

In the \comalow~runs we find very comparable spectra, with only a slightly higher total emission in the case of \comalow.
This difference is however so small that it can also be attributed to differences in the evolution of the cluster between the two simulations.
In general, we find that most spectra are better fit by two power laws, rather than one.
Due to the complex nonlinearities arising from the inclusion of an on-the-fly treatment of CRs in a fully cosmological simulation, no two simulations will be perfectly identical.
This leads to a discrepancy between runs of CR injection both in time and space, which with the short-lived nature of high-energy electrons can lead to variations in the synchrotron spectra.\\
\subsection{Individual CR spectrum}
\begin{figure*}
    \centering
    \includegraphics[width=17cm]{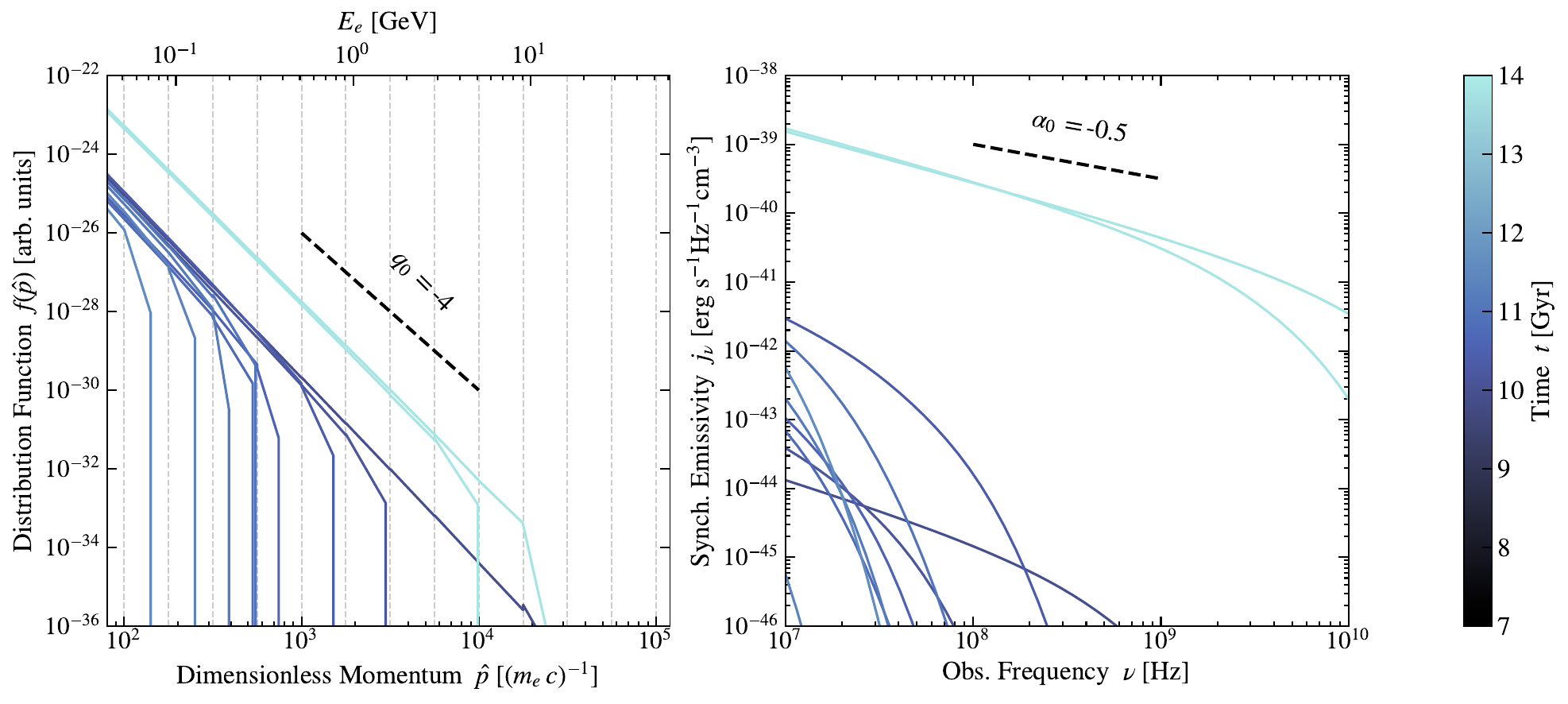}
    \caption{Momentum- and synchrotron spectrum of an arbitrarily chosen SPH particle with an active CR electron population. We show the spectra at every second available output time between $z=1$ and $z=0$ in the \comazoom~simulation.
    Left: Momentum spectrum of the electron population in the momentum range relevant for synchrotron emission. Dashed lines indicate the momentum bin boundaries imposed in our CR scheme. Right: The resulting synchrotron emissivity spectrum of the same electron population.}
    \label{fig:cr_spectrum}
\end{figure*}
To visualize the time evolution of the CR electron spectrum contained in an individual SPH particle we show one such evolution in Fig. \ref{fig:cr_spectrum}.
The SPH particle is taken from the \comazoom~simulation, since it was run with the highest number of output snapshots.
In the left panel, we see the momentum spectra with bin boundaries indicated as dashed lines.

The spectral evolution indicates that the particle has experienced a shock at least twice: Once around 10 Gyrs, and a second time around 14 Gyrs.
This is in agreement with recent results by \citet{Smolinski2023} who find similar multi-shock scenarios can provide the seeds for bright radio relics.
However, none of the output snapshots shown in this figure have a detected shock for the SPH particle, indicating that all acceleration happened between the available output times.
This would lead to an underestimation of synchrotron emission in a pure post-processing approach.
We address this again in Sect. \ref{sec:HB07}.

Between the acceleration episodes, the spectrum is subject to adiabatic expansion, as is evidenced by the left shift of the spectrum at a constant spectral slope.
Energy losses can be seen as the spectral cutoff moving to lower momenta and a steepening of the spectrum below this cutoff.
We evolve this spectral cutoff as an explicit variable, allowing us to represent partially filled momentum bins, as is evident multiple times in the shown spectra.

In the right panel of Fig.~\ref{fig:cr_spectrum}, we show the synchrotron emissivity as a function of frequency for the corresponding CR electron spectra in the left panel.
With readability in mind, we only show the synchrotron spectra based on the simulated magnetic field and omit the other magnetic field models here.

For the synchrotron spectra corresponding to recently injected momentum spectra, we find unbroken power laws up to the 10 GHz shown here, most evident by the spectrum injected at the second acceleration event.
They show slopes close to the theoretical maximum of DSA, consistent with the momentum spectra approaching the same limit.
As the momentum spectra cool from synchrotron and IC cooling we can see a corresponding steepening in the synchrotron spectra.
This primarily affects the high-frequency end of the synchrotron spectra, with the low-frequency end being dominated by adiabatic changes in the gas and variations in magnetic field strength.
%
% full box
\section{Full box}
\label{sec:box}
\subsection{Cosmic ray electron injection}
\begin{figure}
    \centering
    \resizebox{\hsize}{!}{\includegraphics{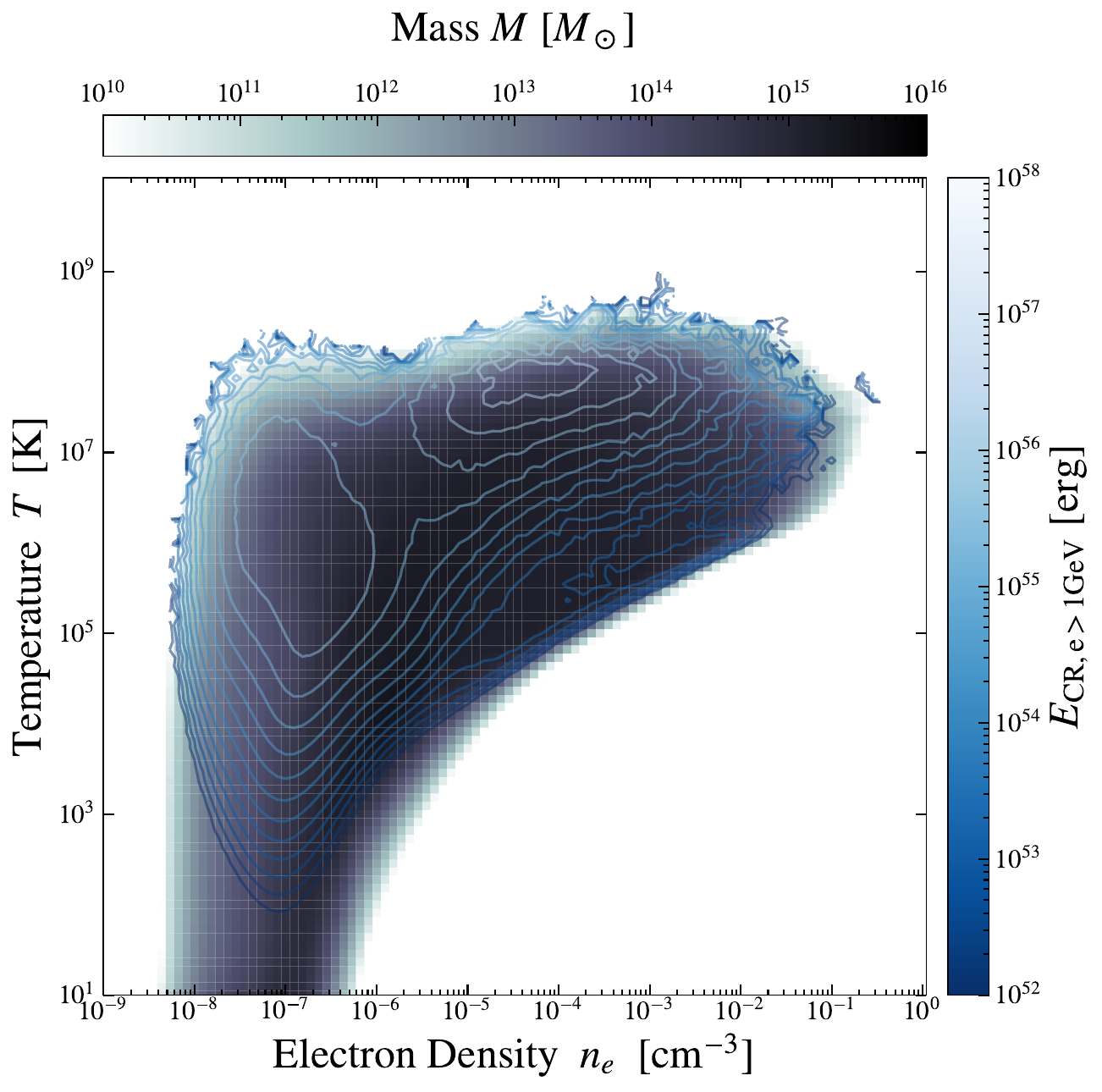}}
    \caption{Density -- temperature phase diagram of all gas particles in the simulation. The colored background shows the mass histogram in phase space. The contours show the CR electron energy of CRs with energies above 1 GeV contained per phase-space element.}
    \label{fig:phase_CReE}
\end{figure}
In Fig.~\ref{fig:phase_CReE}, we show the density-temperature phase-space of all particles contained in \crbox.
The background shows the total mass contained per phase-space element, indicated by the color bar on top.
Contours show the total CR energy per phase-space element, contained in electrons with energies above 1 GeV.
These contours show two main acceleration regions.
One in the moderate density, high-temperature region associated with merger shocks in the cluster periphery.
These CRs could be seen in the form of radio relics, given sufficient magnetic field strength.
The other is in the low density and moderate temperature region of the warm-hot intergalactic medium (WHIM) with temperatures $T_\mathrm{WHIM} \sim 10^5 - 10^7$ K \citep[][]{Cen1999, Dave2001} where we can expect accretion shocks onto clusters and filaments.
This provides a basis for potential synchrotron emission in cosmic web filaments, provided the magnetic field is sufficiently strong.
\subsection{Synchrotron emissivity}
\begin{figure*}[ht]
        \centering
    \includegraphics[width=0.95\textwidth]{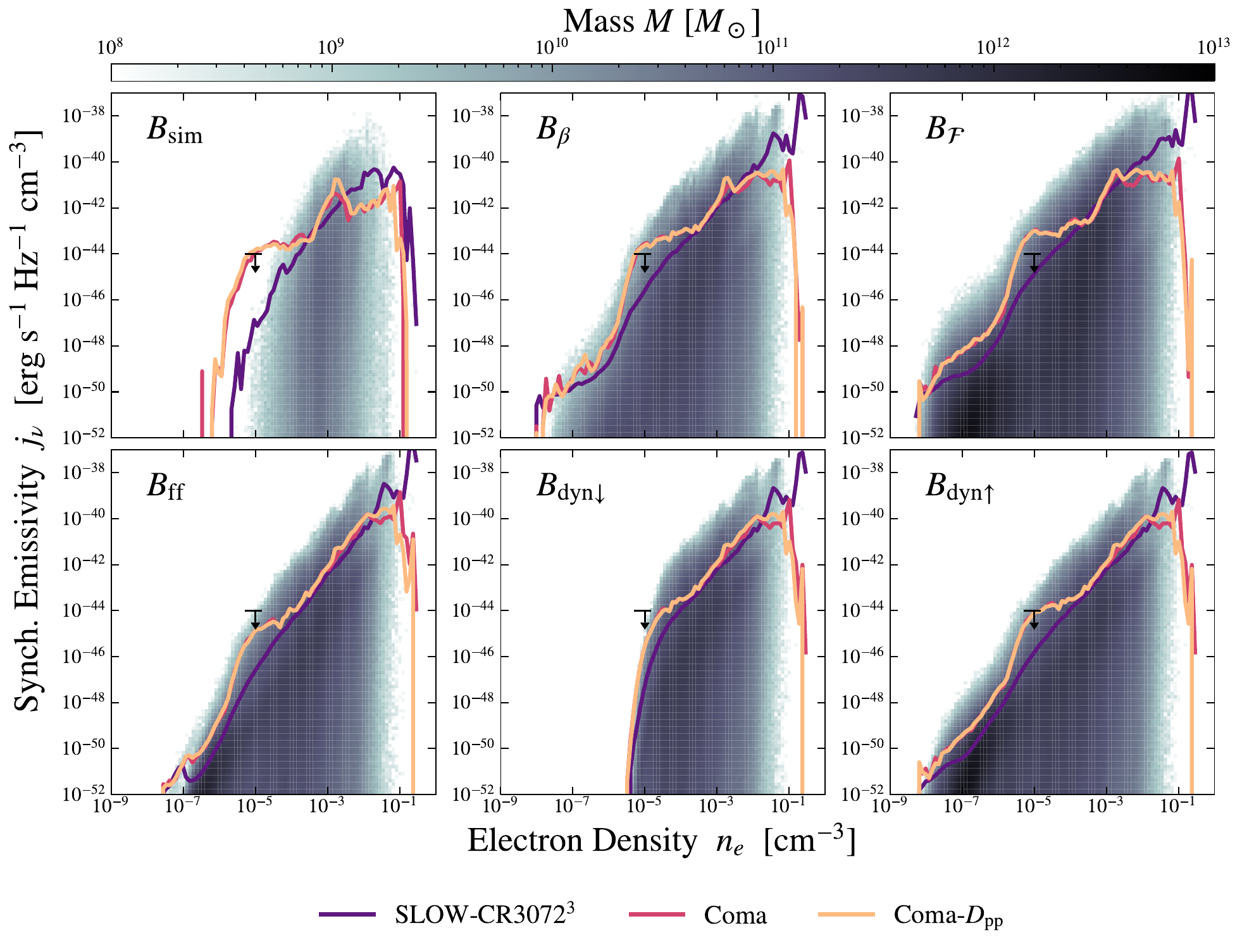}
        \caption{Individual panels showing the phase-space diagrams in the density-synchrotron emissivity phase-space, containing all gas particles in the simulation.
     The background shows mass histograms per phase-space element from the original simulation output.
     The colored lines indicate the mean synchrotron emissivity per density interval for the cosmological box simulation and the zoom-in simulations with and without turbulent reacceleration.
     All emissivities are calculated at a frequency of $\nu = 144$ MHz.
     The upper limit of the synchrotron emissivity is taken from \citet{Hoang2023}.}
    \label{fig:phase_syn}
\end{figure*}
\begin{table}[ht]
    \centering
        \caption{Accompanying Fig.~\ref{fig:phase_syn}, we list the values for the synchrotron emissivity $j_\nu$ at 144 MHz at the $n_e = 10^{-5} \: \mathrm{cm}^{-3}$ bin.}
    \begin{tabular}{l|cccc}
    \toprule
       $B$ Model & \crbox & \comazoom & \comalow \\
        \midrule
       \bsim  & $7.29\times 10^{-48}$ & $1.35\times 10^{-44}$ & $1.72\times 10^{-44}$ \\
       \bbeta & $1.67\times 10^{-45}$ & $1.16\times 10^{-43}$ & $1.01\times 10^{-43}$ \\
       \bturb & $4.35\times 10^{-46}$ & $2.55\times 10^{-44}$ & $2.40\times 10^{-44}$ \\
       \bff   & $3.36\times 10^{-47}$ & $1.53\times 10^{-45}$ & $1.55\times 10^{-45}$ \\
       \bdynl & $1.22\times 10^{-47}$ & $5.52\times 10^{-46}$ & $5.32\times 10^{-46}$ \\
       \bdynh & $1.96\times 10^{-46}$ & $8.94\times 10^{-45}$ & $9.04\times 10^{-45}$ \\
       \bottomrule
    \end{tabular}
    \tablefoot{All values are in units [erg s$^{-1}$ Hz$^{-1}$ cm$^{-3}$]. From left to right we list the different simulations and top to bottom the different magnetic field models introduced in Sect.~\ref{sec:Bfield}.}
    \label{tab:jnu}
\end{table}
Figure~\ref{fig:phase_syn} shows the density-synchrotron emissivity phase space for the simulated and modeled magnetic field.
As in Fig.~\ref{fig:phase_CReE}, the background shows the total mass contained per phase-space element, here for the emissivity calculated from the simulation \crbox.
We only calculated the mean of the particles with emissivities above $j_\nu > 10^{-52}$ erg s$^{-1}$ Hz$^{-1}$ cm$^{-3}$, since the majority of the SPH particles in the simulations do not contain a CRe population, so the mean including all particles is close to zero.
Lines show the mean value per density interval for \crbox and the zoom-in simulations \comazoom~and \comalow.
The arrow indicates the upper limit estimated in \citet{Hoang2023} for $n_e = 10^{-5} \: \mathrm{cm}^{-3}$. 
We list the values for $j_\nu$ we find in this density bin in Table~\ref{tab:jnu}.

In the case of \crbox~, we fall below the upper limit of \citet{Hoang2023} by at least one order of magnitude for all considered magnetic field configurations.
As was discussed in the previous section, this can be attributed to the lack of resolution to capture all of the spurious accretion shocks.
Lower limits in \crbox~can be found with the models \bff and \bdynl, which both show synchrotron emissivities three orders of magnitude below the observational limits.
The more optimistic magnetic field models \bbeta, \bturb~and \bdynh$\:$ still lie significantly below the observational limit by 1.5-2 orders of magnitude.

The zoom-in simulations again show the impact of resolution.
With the shock finder performing better at this resolution the spurious accretion shocks are found more consistently, which leads to a consistent increase in synchrotron emissivity in low-density regions.
This can be seen in the bump of synchrotron emissivity below $n_e \sim 5 \times 10^{-4}$ cm$^{-3}$ which is where we expect to see the turnover from internal merger shocks to accretion shocks \citep[e.g.,][]{Vazza2011}.
The inclusion of turbulent reacceleration does not have a noticeable impact on the emissivity in filaments, at least within the modeling present in this work, as was mentioned above.
Future work will need to account for different reacceleration models to study the potential impact on filaments.

In the case of the simulated magnetic field model \bsim $\:$we find that the mean emissivity in \comazoom~is in agreement with the emissivity limit.
We note, however, that this comparison has its limits since we are comparing mean emissivities between a volume-filling model in \citet{Hoang2023} and spurious shocks in this work.

Compared with the other magnetic field models we find that the increase in resolution in \comazoom~consistently increases the mean synchrotron emissivity by two orders of magnitude compared to \crbox.
As this increase is magnetic field model independent we can conclude that it is injection driven, meaning that we capture more of the spurious accretion shocks in the zoom-in simulation \comazoom, compared to the full cosmological box \crbox.

The lower magnetic field models \bff and \bdynl~both result in emissivities orders of magnitude below the observational limit.
Especially in the case of \bdynl, which matches the 30 nG field inferred by \citet{Carretti2022} at $n_e \approx 10^{-5}$ cm$^{-3}$, this underlines the point of a significant challenge for observations to detect diffuse emission, even by stacking.

For \bbeta~and \bturb~we find emissivities above the observational limit, we note however that this model is biased toward accretion shocks due to its scaling with the thermal and turbulent pressure.
\bdynh~again is roughly in agreement with the observational limit.

As we discuss in Sect. \ref{sec:discussion}, one to two orders of magnitude of synchrotron emissivity can be covered by a variation in the choice of CR injection parameters, giving the possibility of diffuse emission just below the detection limit, even with more conservative magnetic field estimates.

% discussion
\section{Discussion}
\label{sec:discussion}

\subsection{Comparison to post-processing\label{sec:HB07}}
First, we want to compare the result from our on-the-fly solver to the post-processing model by \citet{Hoeft2007}.
We use their acceleration efficiency function $\Psi$ and $\xi_e = 10^{-5}$, similar to \citet{Nuza2017}.
We applied their model to the \comazoom~simulation to make use of the improved shock finder performance compared to \crbox~and show the results in the lowest panels of Fig.~\ref{fig:coma_synch}.

For the relics in our Coma replica, we find that the model by \citet{Hoeft2007} provides reasonable results and most notably, sharper relics.
While in our simulations the region between cluster center and relic is filled with electrons that have been accelerated in the shock passage and subsequently advected downstream of the shock, a post-processing model only accounts for the acceleration at the currently detected shock.
This naturally leads to thinner shocks, as the downstream CRs are not accounted for.

However, when it comes to the diffuse model synchrotron emission of filaments a post-processing model faces a number of limitations.
For one it is naturally coupled to the performance of the shock finder at the time of output.
As can be seen in the panels for \bbeta, \bturb~and \bdynh, which produce volume-filling synchrotron emission with our on-the-fly treatment of CRs, this emission is comparable in absolute flux where it is present; however, it is confined only to the regions of active shock detection.
This leads to very spurious emission only at the detected accretion shocks.
Shock detection in simulations is numerically challenging and requires a number of filtering steps to distinguish between real shocks, cold fronts, and shearing flows \citep[see e.g.,][]{Vazza2009, Schaal2015, Beck2016a}.
This leads to highly transient shock surfaces and introduces difficulties in finding consistent shock surfaces in and around galaxy clusters and filaments at a single timestep.
To add to the complication this naturally scales with numerical resolution, making it easier to detect shocks in the higher resolved intracluster medium (ICM) than in the lower resolved warm-hot intergalactic medium (WHIM) \citep[see e.g.,][for a comparison between different numerical models]{Vazza2011}.

While this is true for every individual timestep of the simulation, we find that our on-the-fly treatment of CRs, and the fact that they can potentially be injected at every timestep in every particle, helps to alleviate this problem.
As a SPH particle moves through the region of an accretion shock, it is highly likely that the shock will be detected at some point through the shock passage and we are not necessarily reliant on the detection happening at the output timestep.
This leads to the smoother injection of CRs and with that a more diffuse emission in the filament volume, as can be seen in the direct comparison between the lower two rows of the rightmost column in Fig.~\ref{fig:coma_synch}.\\
We also find that the advection of CRs with the thermal gas plays a role in the volume-filling synchrotron emission of filaments.
This can be seen in a direct comparison between, for example, the rightmost row for \comazoom, \comalow~and the post-processing approach in Fig.~\ref{fig:coma_synch} where we find more volume-filling emission in the case of the on-the-fly treatment of CRs.
Naturally, a pure post-processing model cannot account for this process and is therefore entirely driven by in situ emission.
This has been addressed by multiple groups and has led to several recent developments of Fokker-Planck solvers with or without tracer particles to account for CR electron injection and transport in large-scale simulations \citep[e.g.,][]{Wittor2017, Yang2017, Winner2019, Ogrodnik2020, Vazza2021, Hopkins2021a}.

\subsection{Choice of cosmic ray injection parameters}
\begin{table*}
    \centering
    \caption{Injection parameters used in this work and selected work in the literature on nonthermal emission by CRs in the large-scale structure of the Universe.}
    \begin{tabular}{|l|c|c|c|c|c|}
    %\toprule
        Reference & $\eta(\mathcal{M}_s)$ & $\eta(\theta_B)$ & $K_\mathrm{ep}$ & $\hat{p}_\mathrm{inj,e}$\\
        \midrule
        This Work & \citet{Ryu2019} & \citet{Pais2018} & 0.01 & 1  \\
        \citet{Hoeft2008} & \citet{Hoeft2007} & - & 0.05 & Eq. \ref{eq:pmin_HB} \\
        \citet{Pinzke2013} & \citet{Pinzke2013} & - & Eq. \ref{eq:Kep} & Eq. \ref{eq:p_inj} \\
        \citet{Hong2014} & \citet{Kang2013} & - & - & 0.01 \\
        \citet{Vazza2015a} & \citet{Kang2013} & 0.3 & 0.01 & Eq.~\ref{eq:pinj_KR13} \\
        \citet{Donnert2016} & \citet{Kang2013} & - & 0.01 & 0.1\\
        \citet{Wittor2017} & \citet{Kang2013} & $\Theta(\theta - \theta_\mathrm{crit})$ & 0.05 & Eq. \ref{eq:pmin_HB}\\
        \citet{Ha2023} & \citet{Ha2023} & - & Eq. \ref{eq:Kep} & Eq. \ref{eq:p_inj} 
    \end{tabular}
    \label{tab:CR_inj}
    \tablefoot{From left to right, we list the indicated work, Mach number-dependent acceleration efficiency, magnetic field angle-dependent efficiency, injection ratio between electrons and protons, and injection momentum.}
\end{table*}
\begin{figure}
    \centering
    \resizebox{\hsize}{!}{\includegraphics{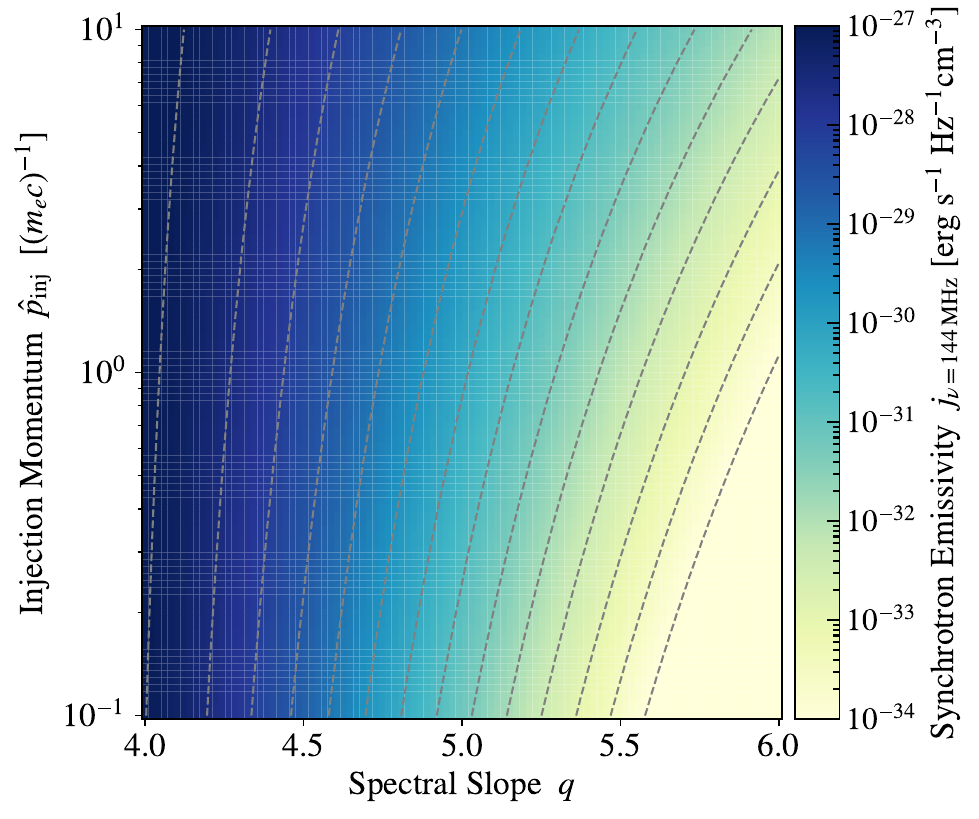}}
    \caption{Synchrotron emissivity of an electron spectrum with a power-law slope, $q$, extending between an injection momentum, $\hat{p}_\mathrm{inj}$, and a maximum momentum of $\hat{p}_\mathrm{max} = 10^6$. Each spectrum is normalized to contain a total energy density of $\epsilon = 1$ erg cm$^{-3}$ and emits in a magnetic field with $B = 1 \mu$G. The color shows emissivity and contour lines are spaced apart by half an order of magnitude.}
    \label{fig:emissivity_p_scaling}
\end{figure}
In this work, we have used a fixed set of CR parameters and evolved the injected CR population over time.
These parameters are, namely, the injection momentum, $\hat{p}_\mathrm{inj}$, as the momentum where the nonthermal power law of the CR electron population starts, the efficiency, $\eta(\mathcal{M}_s, X_\mathrm{cr}, \theta_B)$, as the fraction of shock energy that is available to accelerate CRs to relativistic energies, and finally the ratio between proton and electron injection, $K_\mathrm{ep}$.
Most of these parameters are poorly constrained in the high Mach number, high-$\beta$ plasma shocks expected in and around cosmic web filaments, with only a small number of studies available \citep[e.g.,][]{Kang2013, Guo2014, Ryu2019, Kang2019, Kobzar2021, Ha2023}.
We shall briefly discuss the impact of these parameters on the synchrotron emission in cosmic web filaments.
An overview of our parameters and the ones used in previous work in the literature can be found in Table~\ref{tab:CR_inj}.

\subsubsection{Injection momentum, $\hat{p}_\mathrm{inj}$}
The strongest impact is caused by the choice of the connection point between thermal and nonthermal electron populations, $\hat{p}_\mathrm{inj}$, beyond which the power-law distribution starts.
Typical fixed values used in previous work on this point lie around $\hat{p}_\mathrm{inj} = 0.01 - 0.1$ \citep[e.g.,][]{Hong2014, Donnert2016}.
For the electrons to be efficiently accelerated by DSA, their gyro radius must be on the order of that of the protons, which means a dependence of the injection momentum on the magnetic field and the temperature downstream of the shock.

\citet{Kang2013}, for example, model this as
\begin{equation}
    p_\mathrm{inj} \approx 1.17 m_p v_2 \left( 1 + \frac{1.07}{\epsilon_B} \right)
    \label{eq:pinj_KR13}
,\end{equation}
where $m_p$ is the proton mass, $v_2$ is the downstream velocity of the shock, and $\epsilon_B = B_0/B_\perp$ is the ratio between the mean magnetic field along the shock normal and the amplitude of the post-shock MHD turbulence.

The \citet{Hoeft2007} model omits the explicit magnetic field dependence and relates the injection momentum only to the temperature downstream of the shock ($T_2$).
This leads to an injection momentum of
\begin{equation}
    \hat{p}_\mathrm{inj,e} \sim \frac{10 k_B T_2}{m_e c^2}
    \label{eq:pmin_HB}
,\end{equation}
which for typical shocks in the ICM with $T_2 \sim 10^8$ K is of the order $\hat{p}_\mathrm{inj,e} \sim 0.1-0.2$.\\
A similar parametrization is used, for example, in \citet{Kang2019, Vazza2021, Ha2023} with
\begin{equation}
    \hat{p}_\mathrm{inj,e} \approx \frac{3 \sqrt{\frac{m_e}{m_p}} p_\mathrm{th,p}}{m_e c} = \frac{3 \sqrt{2 k_B m_e T_2}}{m_e c}
    \label{eq:p_inj}
,\end{equation}
which for the shocks in our simulated cosmic web filaments with typical downstream temperatures of $T_2 \sim 10^7$ K would be of the order $\hat{p}_\mathrm{inj} \approx 0.17$.
This makes our choice of a fixed $\hat{p}_\mathrm{inj} = 1.0 $ an optimistic assumption, especially for cold accretion shocks.
We show the impact of the choice of the injection momentum on the synchrotron emissivity in Fig. \ref{fig:emissivity_p_scaling}.
Given the high sonic Mach number, we detect in the surrounding of filaments with $\mathcal{M}_s \gg 10$ the standard RH jump conditions converge toward $q = 4$.
For this value of $q$, even an injection momentum two orders of magnitude higher only introduces a synchrotron emission half an order of magnitude higher.

\subsubsection{Acceleration efficiency, $\eta(\mathcal{M}_s, X_\mathrm{cr}, \theta_B)$}
\begin{figure}
    \centering
    \resizebox{\hsize}{!}{\includegraphics{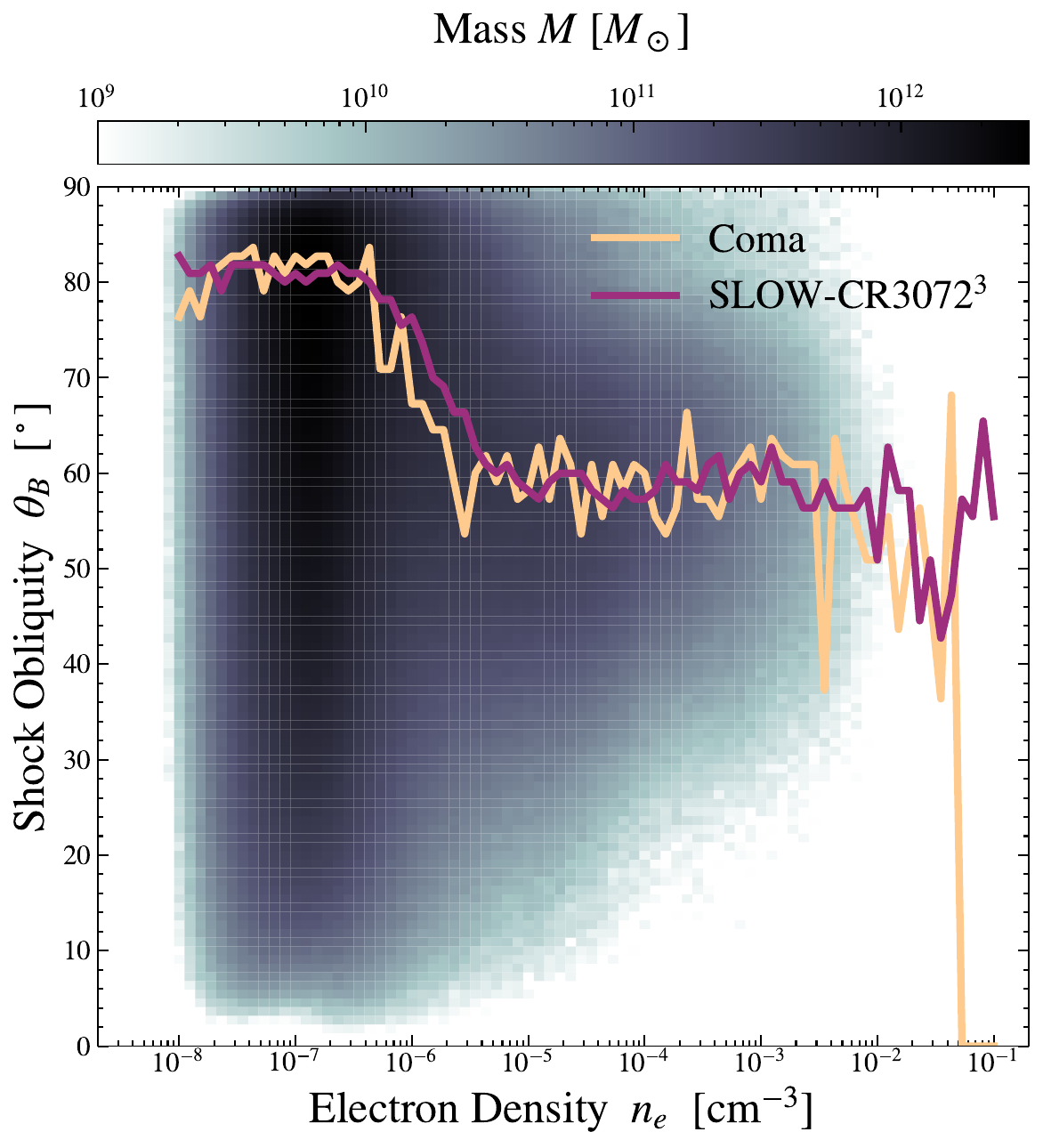}}
    \caption{Density -- shock obliquity phase map. The colors indicate the 2D mass histogram for \crbox, while the lines indicate the maximum of the histograms for \crbox~and \comazoom~as a function of density.}
    \label{fig:phase_theta}
\end{figure}
Mach number-dependent efficiency models that have been used in previous simulations typically show a strong dependence on Mach number for weak shocks $\mathcal{M}_s < 5$ and saturate at a fixed value for strong shocks $\mathcal{M}_s > 7-10$ \citep[e.g.,][]{Hoeft2007, Kang2007, Kang2013, Caprioli2014, Ryu2019}.
This saturation value varies by about one order of magnitude (see the left panel of Fig.~\ref{fig:synch_dsa_models} for our parametrization of the different DSA models).
For accretion shocks, whose Mach numbers lie within the range of the efficiency models above the saturation value this means that the injected energy and with that the synchrotron emission scales linearly with this saturation value.
With our choice of the \citet{Ryu2019} efficiency model, we lie at the lower end of the range for the saturation values.
A different model choice would therefore increase the potential synchrotron brightness by up to a factor of 25.
We note, however, that DSA models for high sonic Mach number, high-$\beta$, and low-temperature shocks are scarce, and further investigation in this direction is needed.

In the case of magnetic field angle dependent acceleration efficiency ($\eta(\theta_B)$), \citet{Vazza2017, Wittor2017} find that allowing CR electron acceleration only at quasi-perpendicular shocks with $\theta > 60^\circ$ does not significantly alter the expected emission from accretion shocks onto clusters and cosmic web filaments.
This stems from the alignment of the shock surface and magnetic field, due to the adiabatic compression and with that amplification of the magnetic field only working on the perpendicular component.
In Fig. \ref{fig:phase_theta}, we show the density--shock obliquity phase space of \crbox~and \comazoom.
We find a similar trend of shock obliquity increasing for low-density regions.

However, we also see a shock obliquity of $\theta_B \approx 60^\circ$ for higher densities.
This result is in agreement with the findings of \citet{Banfi2020} (see their Fig.~17).
\citet{Wittor2017} show that for any upstream shock obliquity, $\theta_\mathrm{pre}$, the shock passage will always align the obliquity toward $\theta = 90^\circ$.
This process only depends on the upstream obliquity and the shock compression ratio, which makes it more efficient for higher Mach number shocks.

Hence, if accounting for a shock obliquity-dependent acceleration efficiency, we preferentially accelerate CR electrons over CR protons.
This can help to ease the tension between CR acceleration efficiency models and the lack of diffuse $\gamma$-ray detection in galaxy clusters \citep[see e.g.,][for a discussion]{Wittor2019}.
We shall address this question as well in upcoming work, analyzing the proton component of our \crbox~simulation.

\subsubsection{Electron-to-proton ratio, $K_\mathrm{ep}$}
The final free parameter is the electron-to-proton injection ratio, $K_\mathrm{ep}$.
This is poorly constrained and is typically taken as a canonical value from SNe as $K_\mathrm{ep} = 0.001 - 0.01$ \citep[e.g.,][]{Park2015}.
\citet{Kang2020} follow an alternative approach by making this parameter injection slope dependent as
\begin{equation}
    K_\mathrm{ep} = \left( \frac{m_e}{m_p} \right)^{(q_\mathrm{inj}-3)/2} \quad .
    \label{eq:Kep}
\end{equation}
At strong accretion shocks with an injection slope of $q_\mathrm{inj} \approx 4$, this gives a value of $K_\mathrm{ep} \approx 0.023$.
This means we underestimate the injection of CRe by a factor of 2.3 by using our constant value.
However, it is not clear what role nonlinear DSA effects can play in these shock environments and if this can lead to a steepening of the injection slope \citep[see][]{Caprioli2020, Diesing2021}.
Given the sensitivity of $K_\mathrm{ep}$ on the injection slope and the uncertainty of the real slope, we note that further study is required.

\citet{Vazza2015} test this value for radio relics and find that to explain observed radio emission they require $K_\mathrm{ep} \geq 10 - 10^{-2}$ to avoid $\gamma$-ray emission in conflict with Fermi-LAT observations.
Recent results indicate that this over-injection of CR protons can also be remedied by the shock-obliquity dependent acceleration of CRs \citep[see][]{Wittor2021}.
Given that the observed brightness of radio relics is still an open problem, this variance in values for $K_\mathrm{ep}$ introduces a large variance in potential synchrotron emission.
The emission in our model scales linearly with $K_\mathrm{ep}$, so larger values for $K_\mathrm{ep}$ can significantly boost the synchrotron emission in filaments.
Recent PIC simulations by \citet{Gupta2024} indicate that electron acceleration may become more efficient compared to proton acceleration at high Mach number shocks, leading to $K_\mathrm{ep} \sim 0.1$.
This would increase the synchrotron emission by an order of magnitude compared to our current modeling.
We shall investigate the impact of different values for $K_\mathrm{ep}$ in future zoom-in simulations.

\subsection{Impact of magnetic field models on synchrotron cooling}
\begin{figure}
    \centering
    \resizebox{\hsize}{!}{\includegraphics{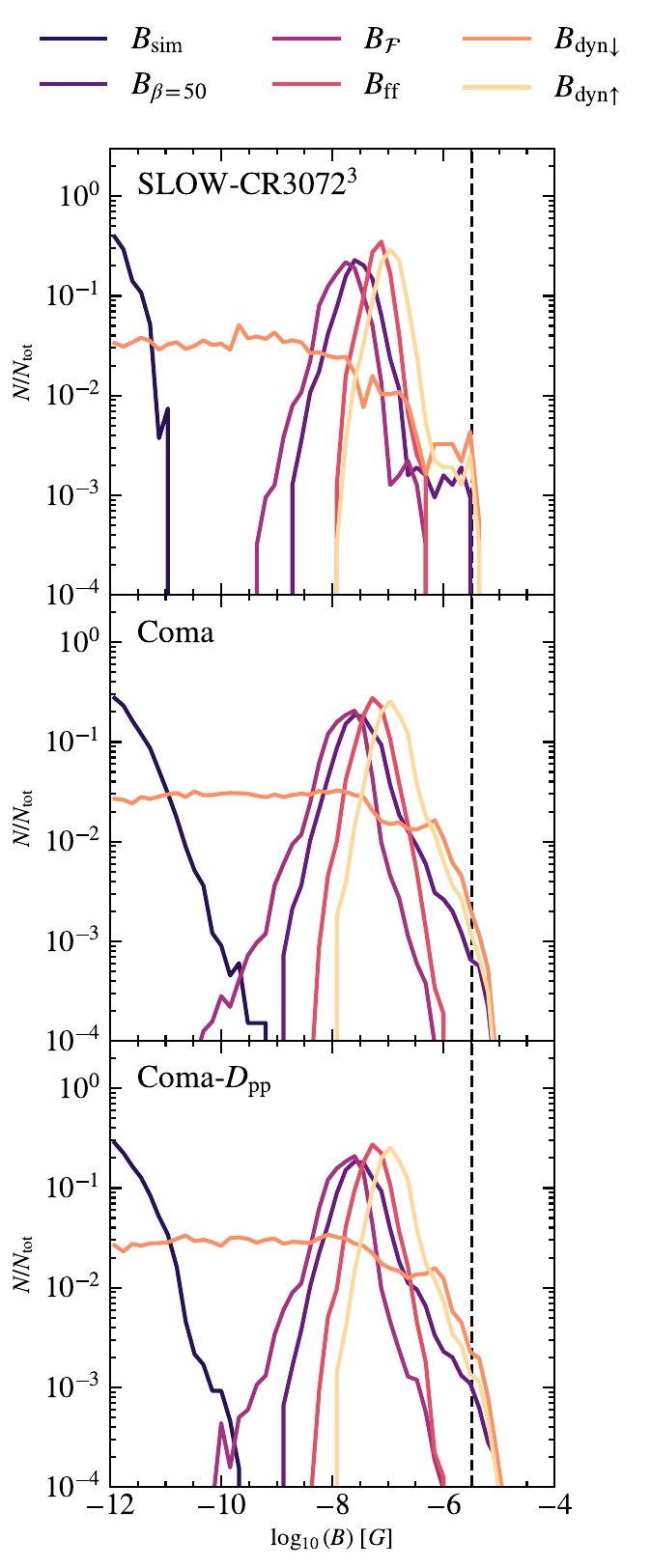}}
    \caption{Histograms of magnetic field strength in all SPH particles containing a CR electron population in the cutout selected for Fig.~\ref{fig:cr_spectrum}. Colored lines correspond to the magnetic field models, the dashed black line shows the equivalence magnetic field of the CMB at $z=0$.}
    \label{fig:B_hist}
\end{figure}
On the point of cooling, specifically cooling due to synchrotron losses, we note that the spectral evolution should in principle be performed again for each of our different magnetic field models.
Since we change the magnetic field model only in post-processing, while keeping the CR spectra the same we can underestimate the cooling due to synchrotron emission that the CR spectra experience in cases where the modeled magnetic field is higher than the simulated one.

However, we find that in the vast majority of the particles containing a CR population cooling is IC scattering dominated since $B < B_\mathrm{CMB} \approx 3.2 \: \mu \mathrm{G} \: (1 + z)^4$.
As can be seen in Fig~\ref{fig:Bfield_radial}, in the example of Coma this concerns all particles beyond the inner $\sim500$ kpc.
Since, for this work, we are primarily interested in the synchrotron emission of cosmic web filaments, whose magnetic field is expected to lie significantly below $B_\mathrm{CMB}$, we aim to discuss the impact of varying magnetic field models on cooling in this section.

In Fig. \ref{fig:B_hist}, we show the histograms of magnetic field strength for all SPH particles contained in the selection used to obtain Fig. \ref{fig:cr_spectrum}.
In addition, we limit the histogram computation to particles that also contain a CR electron population.
The panels correspond to the four simulations used in this work and colors correspond to magnetic field models.
The dashed black line indicates the CMB equivalent magnetic field at $z=0$.
All histograms are normalized to the total number of SPH particles with a CR electron population.

In all cases, the relative number of CR spectra that would experience first-order effects from stronger synchrotron cooling in the case of stronger magnetic field models is of the order $N/N_\mathrm{tot} \propto \mathcal{O}(10^{-3})$.
We further note that for these particles to have such high magnetic fields in our models they have to lie close to the center of the filament and should hence have undergone significant cooling due to IC scattering since their acceleration at the accretion shock and subsequent advection into the filament.
This reduces the significance of the additional cooling due to synchrotron emission. We therefore do not expect a significant impact on our results from not rerunning the spectral evolution with the various magnetic field models.

\subsection{Impact of AGN feedback and star formation}
With our simulation being non-radiative, we do not account for AGN feedback or star formation.
We therefore also cannot account for CR injection by these processes.
In the following, we briefly discuss the potential impact of this on our results.
An AGN can potentially provide a significant CR electron seed population that can advect and diffuse in the cluster volume \citep[see e.g.,][]{ZuHone2021, Vazza2023, Dominguez2024}.
However, this advection is expected to be confined to clusters, where the most powerful AGNs are expected to be located, not filaments, which are the primary focus of this work.
There the injected electron can potentially be reenergized by cluster merger shocks and can be observed as the so-called radio phoenices \citep[][]{Weeren2019}.
The jets of AGNs can also inject significant turbulence, as for example is observed in Perseus \citep[][]{Sanders2016}.
This turbulence can then reaccelerate fossil electrons and provide the basis for radio halos.

We aim to include AGN feedback in the near future, as the modeling of radio galaxies in our sample should help to estimate the background pollution by discreet sources as well as additional feedback on the cluster gas by AGN-driven shocks and their impact on our results.
This is, however, beyond the scope of this work.

Cosmic ray proton injection at SNe has been found to drive and/or modify the structure of galactic winds and the circumgalactic medium (CGM) \citep[e.g.,][]{Pfrommer2017, Butsky2018, Fielding2020, Ji2020, Girichidis2023}.
Cosmic ray protons generally have lifetimes exceeding the Hubble time in the low-density CGM and ICM \citep[see e.g., Fig 7 in][for an illustration of CR proton cooling times at different densities]{Brunetti2007}, which under the assumption of efficient transport can provide a volume-filling CR proton population.
There they can nonetheless interact with the thermal background gas and be scattered into $\pi$-ons and, depending on the charge of the $\pi$-on, decay into $\gamma$-photons, or electrons or positrons.
These electrons could then contribute to diffuse synchrotron emission.
In the context of CR proton acceleration at merger shocks, this process has been proposed as a source of radio halos \citep[e.g.,][]{Dolag2000}.
However, the non-detection of diffuse $\gamma$-ray emission from galaxy clusters puts a strong upper limit on this process \citep[e.g.,][]{Pinzke2010, Pinzke2011, Ackermann2014}.
For this same reason, we omitted including an explicit treatment of secondary electrons.
\subsection{Observational prospects}
\begin{figure*}
    \centering
    \includegraphics[width=17cm]{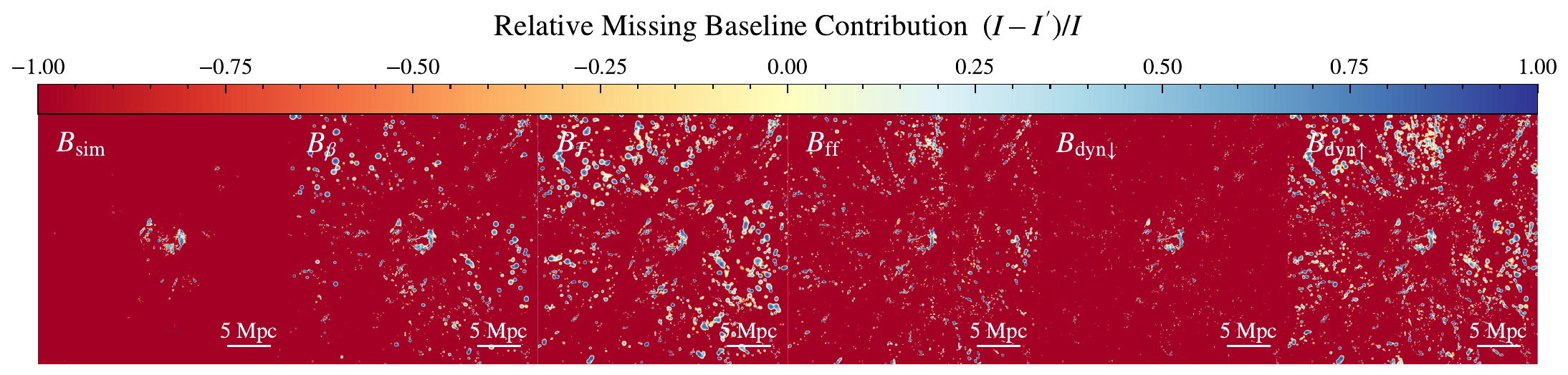}

    \includegraphics[width=17cm]{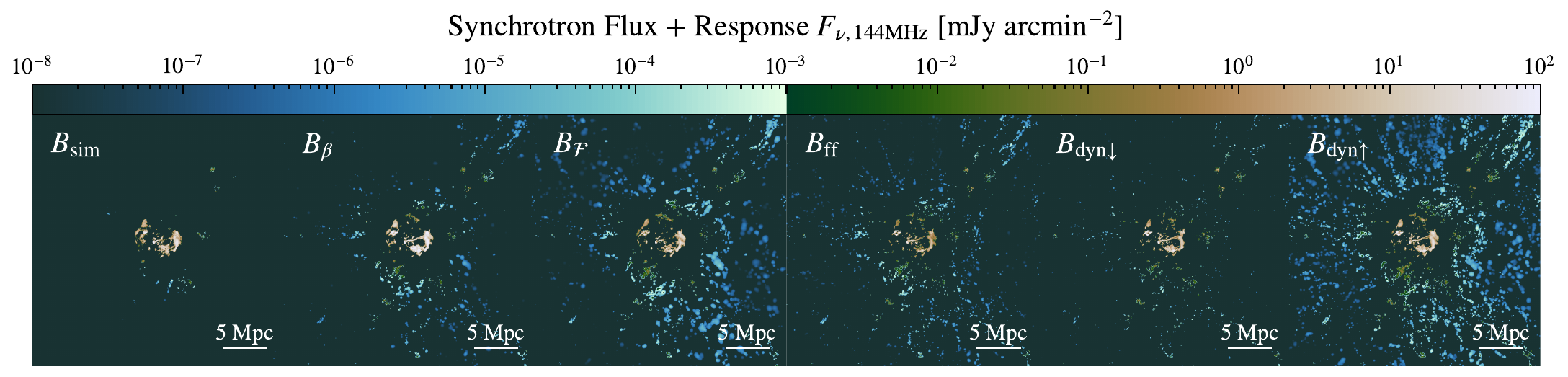}
    \caption{To illustrate the impact of missing baselines on our simulated flux images, mock observations of \comazoom~as in Fig.~\ref{fig:coma_synch} under these considerations. Top: Relative change to the flux due to missing baselines in LOFAR. Bottom: Synchrotron flux under the consideration of instrument response from missing baselines.}
    \label{fig:baselines}
\end{figure*}
Given the potential impact of so-far poorly constrained CR injection parameters on our results, we can make a best-case assumption of the observational requirements for direct observation of diffuse synchrotron emission from cosmic web filaments. 
We shall discuss this again for the filaments we find around our Coma cluster replica, as it provides the widest range of parameters in our set due to the zoom-in re-simulation.
As is shown in Fig. \ref{fig:coma_synch}, the most optimistic magnetic field models, normalized to the observed synchrotron flux from the Coma relic in the higher resolution simulations leads to a diffuse synchrotron flux from the filaments of $S_{\nu = 144 \: \mathrm{MHz}} \approx 0.1 \mu$Jy beam$^{-1}$ for a beam size of $\theta = 60" \times 60"$.
This is three orders of magnitude below the noise of $\sigma_\mathrm{rms} = 0.4$ mJy beam$^{-1}$ given in \citet{Bonafede2022}.
However, the choice of other parameters for $p_\mathrm{inj}$, $\eta(\mathcal{M}_s)$ and $K_\mathrm{ep}$ could boost this emission by up to two orders of magnitude.
This would indicate diffuse emission just a factor of ten below the current noise level. The stacking approach by \citet{Hoang2023} reduces the noise to $\sigma_\mathrm{noise} = 0.6 \: \mu$Jy beam$^{-1}$, which lies just a factor of $\sim$ten above the diffuse emission we find in the most optimistic scenario.
\\
On top of surface brightness limitations, we further need to account for a suppression of observability, due to the large structures we consider and missing baselines of LOFAR.
These missing baselines suppress the observed flux from synchrotron emission on large scales and increase it on small scales.
To estimate the impact of missing baselines on the observability, we can follow the approach of, for example, \citet{Churazov2022}, using Eq. 3.28 in \citet{Remijan2019}.
This gives an expression for the scale $\theta_\mathrm{MRS}$ at which the flux at a given frequency, $\nu_\mathrm{obs}$, is suppressed by the minimum baseline, $D_\mathrm{min}$, as
$\theta_\mathrm{MRS} \approx 0.6 c/(\nu D_\mathrm{min})$.
Using a minimum baseline of $D_\mathrm{min} = 0.1$ km from \citet{Bonafede2021}, we obtain $\theta_\mathrm{MRS} \approx 42.94' \approx 1.24$ Mpc for our Coma replica.\\

To calculate the contribution to the emission, we folded our non-smoothed images, $I$, of Fig.~\ref{fig:coma_synch} with a Gaussian beam of size $\theta_\mathrm{MRS}$ to obtain an image, $I'$.
This gives a relative contribution $(I - I')/I$ and leads to more filamentary emission in the cluster periphery while suppressing diffuse emission in the cluster and filament centers.

The relative contribution can be seen in the top panels of Fig.~\ref{fig:baselines} in the case of \comazoom.
We chose this simulation as an example since it showed the strongest synchrotron emission in the filaments.
To obtain a more realistic instrument response for LOFAR we added the missing baseline contribution to the original flux images and show the result in the bottom panels of Fig.~\ref{fig:baselines}.

As was mentioned above, this addition has two notable effects.
First, it strongly suppressed the diffuse flux from the central parts of filaments, and to a lesser extent the downstream regions of our relic analogs.
Second, it increases flux originating from the collision between accretion and merger shocks we find close to the virial radius.
These shock structures are morphologically similar to the accretion relic candidate reported in \citet{Bonafede2022}; however, in our simulation we find that the emission to the upper right and lower left that most closely resembles the accretion relic originates from a previous, small merger along this axis.
A detailed study of the origin of a possible accretion relic is however beyond the scope of this work. The inclusion of missing baselines in the consideration of observational prospects further reduces the chance of observing this emission, at least in nearby filaments.
%
% conclusions
\section{Conclusions\label{sec:conclusions}}

We have presented results from the first constrained cosmological simulation with an on-the-fly Fokker-Planck solver to study the time evolution of CR electron populations on a cosmological scale.
We used this simulation to study potential diffuse synchrotron emission from cosmic web filaments.
Our results can be summarized as follows:
\begin{itemize}
    \item We find that the acceleration of CR electrons occurs within clusters as well as in accretion shocks onto clusters and cosmic web filaments. The electrons with the longest lifetimes settle into large halos around clusters and in cosmic web filaments. Given enough magnetic field strength and stabilization against cooling losses, by for example Fermi-II reacceleration, these electrons can provide the basis for synchrotron emission in these regions.
    \item We tested six different magnetic field configurations to study the threshold for obtaining considerable diffuse emission from cosmic web filaments and found that with our choice of CR injection parameters we require fields of the order $B \sim 100 $ nG to obtain considerable emission.
    \item To test the impact of resolution in our simulation and the inclusion of on-the-fly turbulent reacceleration, we performed zoom-in simulations of our Coma cluster replica. We find that higher resolution, and with that a more accurate capture of accretion shocks, increases the synchrotron emission by an order of magnitude in low-density regions. This would in turn decrease the required magnetic field strength to obtain the same level of synchrotron emission as in the low-resolution full box by three orders of magnitude, bringing it into agreement with current observations with $B \approx 30$ nG.
    \item The inclusion of turbulent reacceleration in our modeling does not increase the synchrotron emission in filaments, at least in the gyro-resonant flavor used for this work.
    \item With our chosen parameters for CR injection, the surface brightness of diffuse emission from cosmic web filaments lies three orders of magnitude below the current noise limit for direct detection with \textsc{LOFAR}.
    \item In the most optimistic model for magnetic field strength and CR injection parameters, we can estimate synchrotron emission only a factor $\sim 5$ below the noise level currently achievable by stacking of radio images.
    \item In the models that provide significant synchrotron emission from the filaments, we find that the spectral index of diffuse emission in the LOFAR band is in the range of $\alpha \approx$ -1.0 -- -1.5, in agreement with recent observations by \citet{Vernstrom2023}.
\end{itemize}
In future work, we shall present the results of \crbox $\:$ for the radio halos of our Coma, Perseus, and Virgo replicas, radio halo relations, and $\gamma$-ray emission from the CR proton population.
We shall further increase our sample size of zoom-in simulations and vary the parameters for CR injection and turbulent reacceleration to study their impact on the time evolution of radio halos.

\begin{acknowledgements}
We want to thank the anonymous referee for helpful comments that improved the quality of this work.
We thank Sebastian Nuza for providing us with the fit function for the \citet{Hoeft2007} model.
LMB would like to thank Kamlesh Rajpurohit, Jennifer Schober, Mathias Hoeft, Sebastian Nuza and Damiano Caprioli for helpful discussions.
LMB, KD and IK acknowledge support by the Deutsche Forschungsgemeinschaft (DFG, German Research Foundation) under Germany's Excellence Strategy – EXC 2094 – 390783311 as well as support for the COMPLEX project from the European Research Council (ERC) under the European Union’s Horizon 2020 research and innovation program grant agreement ERC-2019-AdG 882679.
KD, EHM, BS and JS acknowledge support by the grant agreements ANR-21-CE31-0019 / 490702358 from the French Agence Nationale de la Recherche / DFG for the LOCALIZATION project.
UPS is supported by the Simons Foundation through a Flatiron Research Fellowship at the Center for Computational Astrophysics of the Flatiron Institute. The Flatiron Institute is supported by the Simons Foundation.
The calculations for the hydro-dynamical simulations were carried out at the Leibniz Supercomputer Center (LRZ) under the project pn68na.

\end{acknowledgements}

%%%%%%%%%%%%%%%%%%%% REFERENCES %%%%%%%%%%%%%%%%%%

% The best way to enter references is to use BibTeX:

\bibliographystyle{aa}
\bibliography{example} % if your bibtex file is called example.bib

% Alternatively you could enter them by hand, like this:
% This method is tedious and prone to error if you have lots of references
%\begin{thebibliography}{99}
%\bibitem[\protect\citeauthoryear{Author}{2012}]{Author2012}
%Author A.~N., 2013, Journal of Improbable Astronomy, 1, 1
%\bibitem[\protect\citeauthoryear{Others}{2013}]{Others2013}
%Others S., 2012, Journal of Interesting Stuff, 17, 198
%\end{thebibliography}

%%%%%%%%%%%%%%%%%%%%%%%%%%%%%%%%%%%%%%%%%%%%%%%%%%

%%%%%%%%%%%%%%%%% APPENDICES %%%%%%%%%%%%%%%%%%%%%

\appendix

\section{Acceleration models \label{app:eta}}
\begin{figure*}
    \centering
    \includegraphics[width=17cm]{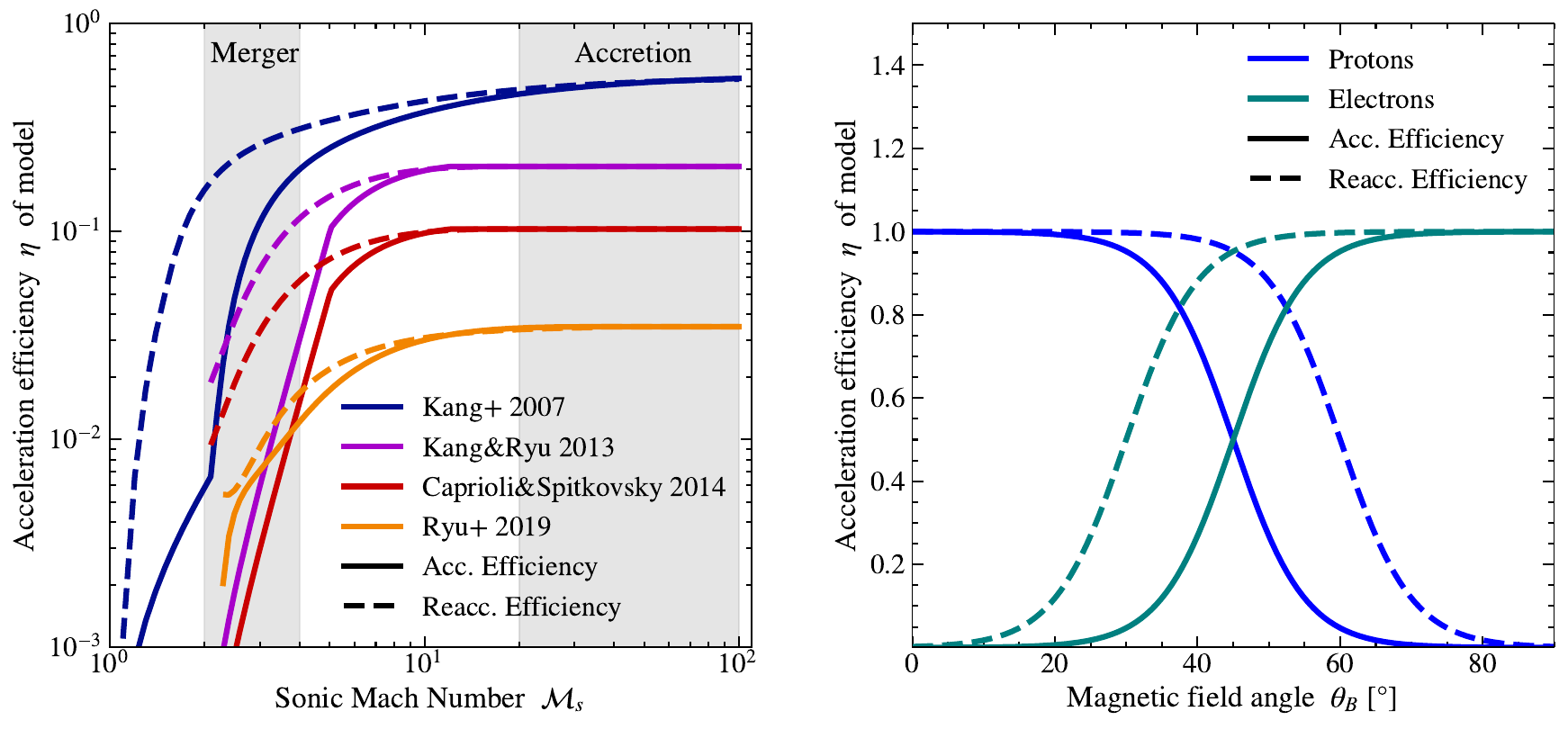}
    \caption{DSA parametrizations present in our model. Left: Sonic Mach number $\mathcal{M}_s$ dependent acceleration models. Right: Shock obliquity ($\theta_B$) dependent acceleration model.}
    \label{fig:synch_dsa_models}
\end{figure*}
We show the DSA parametrizations present in \textsc{Crescendo} in Fig.~\ref{fig:synch_dsa_models}.
For this work, we used the sonic Mach number-dependent acceleration model by \citet{Ryu2019}, which is shown as the orange line in the left panel.
Acceleration efficiency from the thermal pool is shown in solid lines and reacceleration efficiency in dashed lines.
We indicate typical ranges of sonic Mach numbers for merger- and accretion shocks with gray bands.

The right panel shows the shock obliquity-dependent acceleration model.
For protons, we use the parametrization by \citet{Pais2018}, who fit the data by \citet{Caprioli2014}.
The study in \citet{Caprioli2014} was performed for protons, which are found to be preferentially accelerated at quasi-parallel shocks.
Electrons are found to be preferentially accelerated by quasi-perpendicular shocks, so we take the simple approach of shifting the \citet{Pais2018} model by $90^\circ$.
Further studies of the exact dependence of electron acceleration on $\theta_B$ are needed to improve this simple model.

\section{Time integration of the CR spectra}

We shall briefly recap the time integration of \textsc{Crescendo}, to provide the relevant equations for the next section (for details, see \citet{Boess2023}). To update the distribution of the CRs in the two-moment approach we need to compute number- and energy changes per bin over a timestep.
We can obtain the number of CRs per bin by performing a volume-integral in momentum space in shells of the with of the individual bins.
\begin{align}
        N_i = \frac{1}{\rho} \int\limits_{p_i}^{p_{i+1}} dp \: 4 \pi p^2 f(p)
                = \frac{4 \pi f_i p_i^3}{\rho } \frac{\left( \left( \frac{p_{i+1}}{p_i}\right)^{3-q_i} - 1 \right)}{3 - q_i } \label{eq:n_i_analytic}
\end{align}
We calculated the energy contained in each bin in the ultra-relativistic limit $T(p) \approx pc$ and attached this energy to every CR in our previous volume-integral to obtain
\begin{align}
        E_i = \frac{1}{\rho} \int\limits_{p_i}^{p_{i+1}} dp \: 4 \pi c p^3 f(p) 
                = \frac{4 \pi c f_i p_i^4}{\rho} \frac{\left( \left( \frac{p_{i+1}}{p_i}\right)^{4-q_i} - 1 \right)}{4 - q_i } 
  \label{eq:e_i_analytic}
.\end{align}
For the time integration, Eq.~\ref{eq:fp-adiabatic} was rewritten in terms of number- and energy changes per bin $i$ as
\begin{equation}
        \frac{D N_i}{D t} = \frac{1}{\rho} \left[ \left( \frac{1}{3} \frac{\partial u}{\partial x} p + b_l(p) \right) 4 \pi p^2 f(p) \right]_{p_i}^{p_{i+1}}
\end{equation}
\begin{align}
        \frac{D E_i}{D t} =& \frac{1}{\rho} \left[ \left( \frac{1}{3} \frac{\partial u}{\partial x} p + b_l(p) \right) 4 \pi c p^3 f(p) \right]_{p_i}^{p_{i+1}} \\&- \left( \frac{4}{3} \frac{\partial u}{\partial x} E_i + \frac{1}{\rho} \int\limits_{p_i}^{p_{i+1}}  dp \: b_l(p) 4 \pi c p^2 f(p) \right)
\end{align}
 where we introduced the quantity $R_i(q_i, p_i)$ for the energy loss integral per bin
\begin{equation}
        R_i(q_i, p_i) = \frac{4 - q_i}{p_{i+1}^{4-q_i} - p_{i}^{4-q_i}} \int\limits_{p_i}^{p_{i+1}}  dp \:\: p^{2 - q_i} \left( \frac{1}{3} \frac{\partial u }{\partial x} + \sum_{l}^{N_\mathrm{losses}} b_l(p) \right)
        \label{eq:R_losses}
\end{equation}
for simplification.
Integrating this in times gives
\begin{align}
    N_i^{t+\Delta t} &= N_i^t + \frac{1}{\bar{\rho}} \left(F_{N_{i+1}}^m - F_{N_i}^m \right)\\
        E_i^{t + \Delta t} &\approx E_i^t \left( 1 - R_i(q_i, p_i) \right) + \frac{1}{\bar{\rho}} \left( F_{E_{i+1}}^m - F_{E_i}^m \right)
        \label{eq:e_n_update}
\end{align}
where we can associate
\begin{align}
        F_{N_i}^m &= \int\limits_{t}^{t+\Delta t} dt' \: b_l(p) \: 4\pi p^2 f(t', p)\vert_{p_i} \\ 
        F_{E_i}^m &= \int\limits_{t}^{t+\Delta t} dt' \: b_l(p) \: 4\pi c p^3 f(t', p)\vert_{p_i}
        \label{eq:flux_of_t}
\end{align}
with the time-averaged fluxes over the momentum bin boundaries.
We then transfer the problem into momentum space by substituting $dt'$ with the momentum change over time.
This gives equations for the fluxes that depend only on the momentum of a particle $p_u$ at which it must have started out before a timestep $\Delta t$ to arrive at $p_i$.
\begin{align}
        F_{N_i}^m &=  \int\limits_{p_i}^{p_u}  \: dp \: 4 \pi p^2 f^m(p) \label{eq:n_flux_integral}  \\
        F_{E_i}^m &=  \int\limits_{p_i}^{p_u}  \: dp \: 4 \pi c p^3 f^m(p), \label{eq:e_flux_integral} 
\end{align}
with 
\begin{align}
        f^m(p) &= \begin{cases} f_i \left(\frac{p}{p_i}\right)^{-q_i} \quad & \mathrm{if } p_u > p_i \\
                                                        f_{i-1} \left(\frac{p}{p_{i-1}}\right)^{-q_{i-1}} \quad & \mathrm{if } p_u \leq p_i
                          \end{cases} 
\end{align}
After the time integration is finished we end up with new values for CR number and energy per bin.
To reconstruct the distribution function we solve
\begin{align}
        \frac{E_i}{N_i p_{i} c} = \frac{3 - q_i}{4 - q_i} \frac{\left( \frac{p_{i+1}}{p_i} \right)^{4-q_i}-1}{\left( \frac{p_{i+1}}{p_i} \right)^{3-q_i}-1} 
        \label{eq:slope_solver}
\end{align}
for $q_i$ numerically using Brent's method and obtain the updated norm by solving Eq.~\ref{eq:n_i_analytic} for $f_i$ as 
\begin{align}
        f_i = \frac{\rho \: N_i}{4 \pi p_i^3} \frac{3 - q_i}{\left( \frac{p_{i+1}}{p_i} \right)^{3 - q_i} - 1}
        \label{eq:f_i_update}
\end{align}

\section{Fermi-II reacceleration\label{app:Dpp}}

As was discussed above, to model the turbulent reacceleration of CRs we adopt the model by \citet{Cassano2005}\footnote{This adaptation has been implemented into the Gadget3 code-family by Julius Donnert, but to the best of our knowledge has not been published in the presented form.}.
We only account for the systematic component of reacceleration, which can be expressed in the unified-cooling approach, extending the $b_l(p)$ term in Eq.~\ref{eq:fp-rad} to include the momentum change due to systematic reacceleration
\begin{equation}
    \sum_l b_l(p) \equiv \left( \frac{\mathrm{d}p}{\mathrm{d}t} \right)_{\mathrm{cool}}^{\mathrm{synch}} + \left( \frac{\mathrm{d}p}{\mathrm{d}t} \right)_{\mathrm{cool}}^{\mathrm{IC}} + \left( \frac{\mathrm{d}p}{\mathrm{d}t} \right)_{\mathrm{acc}}^{\mathrm{sys}}  .
\end{equation}
This momentum change over time can be expressed as
\begin{equation}
        \left( \frac{\mathrm{d}p}{\mathrm{d}t} \right)_{\mathrm{acc}}^{\mathrm{sys}} = -\chi p \approx - 2 \frac{D_{\mathrm{pp}}}{p}.
        \label{eq:dpdt_reacc}
\end{equation}
For an isotropic distribution of wave and particle momenta and $v_\mathrm{A} < v_\mathrm{M}$, with $v_\mathrm{A}$ being the Alfvén velocity and 
\begin{equation}
        v_\mathrm{M}^2 \approx \frac{4}{3} v_\mathrm{ion}^2 + v_\mathrm{A}^2
        \label{eq:vM}
\end{equation}
the momentum diffusion coefficient, $D_{\mathrm{pp}}$, is given by \citet{Eilek1979} as 
\begin{align}
        D_{\mathrm{pp}}(p, t) &\approx 4.45 \pi^2 \frac{v_\mathrm{M}^2}{c} \frac{p^2}{B^2} \int\limits_{k_\mathrm{min}}^{k_\mathrm{max}} \mathrm{d}k \: k\mathcal{W}_k^B(t) 
                                \label{eq:Dpp_integral}\\
                                 &= D_0(t) \: p^2 \quad .
        \label{eq:Dpp}
\end{align}
This assumes that the infall of substructure causes plasma instabilities and turbulence which in turn drives magneto-sonic (MS) waves.
These instabilities lead to the systematic second-order Fermi reacceleration process.
The integral on the r.h.s. of Eq.~\ref{eq:Dpp_integral} is therefore the integral over the spectrum of the MS waves.
The boundaries of this integral are defined by the minimum and maximum wave numbers of the underlying turbulence spectrum. 
Combining Eq.~\ref{eq:dpdt_reacc} and Eq.~\ref{eq:Dpp} gives the short form of
\begin{equation}
        \left( \frac{\mathrm{d}p}{\mathrm{d}t} \right)_{\mathrm{acc}}^{\mathrm{sys}} = - 2 D_0(t) \: p \: .
        \label{eq:dpdt_Dpp}
\end{equation}
Explicitly solving $D_0(t)$ proves difficult due to the integral on the r.h.s. of Eq.~\ref{eq:Dpp_integral}.
The time evolution of $W_k(t)$, the modified spectrum of the MS waves, depends on wave-wave interaction and wave-particle interaction and is therefore very expensive to solve.
\citet{Cassano2005} use an approximation that simplifies the form of $W_k$ to only depend on the injection spectrum of the waves $I(k)$, which for this purpose is assumed to be a single power law, $I(k) = I_0 k^{-a}$, and the most prominent dampening mechanism, the dampening by thermal electrons, $\Gamma_{\text{th},e}(k)$,
\begin{align}
        W_k \simeq \frac{I(k)}{\Gamma_{\mathrm{th},e}(k)},
\end{align}
with 
\begin{align}
        \Gamma_{\mathrm{th},e}(k) &= \sqrt{32\pi^3} \rho \sqrt{m_\mathrm{e} k_\mathrm{B} T} \left(\frac{v_\mathrm{M}}{B}\right)^2 \frac{W_k^B}{W_k} \mathcal{I}(x) k, 
\end{align}
and
\begin{align}
        \mathcal{I}(x) = 2 \int\limits_1^{\infty} dx \left( \frac{1}{x} - \frac{1}{x^3} \right) \exp \left[ -x^2 \left(\frac{v_M}{v_\mathrm{th}} \right)^2 \right] .
\end{align}
Here $v_M^2$ is given by Eq.~\ref{eq:vM} and $v_\mathrm{th}^2 = \frac{2 k_B T}{m_e}$ is the thermal velocity of the electrons \citep[see Sect. 4.2 of][for more details]{Cassano2005}.
To explicitly solve $D_0$ at every timestep, we express it as 
\begin{align}
        D_0(t) = \frac{4.45 \pi^2}{c \: \sqrt{32 \pi^3 m_\mathrm{e} k_\mathrm{B}}} \:  \frac{\eta_t a_k  E_{\mathrm{turb}}}{n_\mathrm{e} V_\mathrm{p} \: \sqrt{T} \: \Delta t \: \mathcal{I}(x)},
 \label{eq:D0}
\end{align}
where $\eta_t$ is the free parameter introduced by \citet{Cassano2005} with $\eta_t = 0.2$ in this model, $E_\mathrm{turb}$ is the turbulent energy, $V_\mathrm{p}$ is the kernel volume of a particle and $a_k$ is the turbulent scale factor
\begin{align}
        a_k = \frac{k_{\mathrm{max}}^{-2/3} - k_{\mathrm{min}}^{-2/3}}{k_{\mathrm{mps}}^{-2/3} - k_\mathrm{h}^{-2/3}} \quad .
\end{align}
The different components of this are $k_{\mathrm{min}}$ and $k_{\mathrm{max}}$ as the integral limits for Eq.~\ref{eq:Dpp_integral}, the wavelength of the mean particle separation $k_{\mathrm{mps}} = N_{\mathrm{ngb},i}^{1/3}/(2h_i)$ and the maximum wavelength within a kernel of smoothing length $h_i$, $k_\mathrm{h} = 1/(2 h_i)$.
$k_{\mathrm{min}}$ and $k_{\mathrm{max}}$ are free parameters in this model and can be expressed as the inverse of the injection and damping scale of the MS waves which for the purpose of this work take the values
\begin{align*}
        k_{\mathrm{min}} &= ( \Lambda_{\mathrm{inj}} )^{-1} = ( 200 \mathrm{ kpc} )^{-1} \\
        k_{\mathrm{max}} &= ( \Lambda_{\mathrm{damp}} )^{-1} = ( 0.1 \mathrm{ kpc} )^{-1} \quad .
\end{align*}
To estimate the turbulent velocity and from that the turbulent energy we use the same approach as \citet{Donnert2014}.
We calculated $E_\mathrm{turb}$ from the root-mean-square (RMS) velocity which we can estimate by calculating the bulk velocity in the kernel; for example, for particle $i$,
\begin{equation}
    v_{\mathrm{bulk},i} = \sum\limits_{j=0}^{N_\mathrm{ngb}} \frac{m_j}{\rho_j} \mathbf{v}_j W(\vert\mathbf{r_i} - \mathrm{r_j}\vert, h_i)
\end{equation}
and subtracting that from the velocity of particle $i$
\begin{equation}
    \mathbf{v}_{\mathrm{rms},i} = \mathbf{v}_i - v_{\mathrm{bulk},i}
\end{equation}
such that
\begin{equation}
    E_\mathrm{turb} = \frac{1}{2} m \: \vert \mathbf{v}_\mathrm{rms} \vert^2 \: .
\label{eq:Eturb}
\end{equation}
We find that this provides a reasonable proxy for the turbulent energy captured in SPH, even when compared to an estimate based on a full Helmholtz decomposition (Groth et al., in prep).
This approach does however use the total available turbulent energy to reaccelerate electrons.
As was mentioned in Sect.~\ref{sec:fermi2}, recent work by \citet{Brunetti2016, Brunetti2020} suggests that only the solenoidal component of the turbulent velocity can efficiently reaccelerate CRs, therefore our current model will need to be revised to test the impact of these results on our simulations.\\
To evolve our electron spectra we then transform the problem into momentum space as for the energy losses and adiabatic changes.

With Eq.~\ref{eq:dpdt_Dpp} this leads to an upper integration boundary of
\begin{equation}
   p_u = p_i \cdot e^{-2 D_0 \Delta t}
\end{equation}
and a spectral cut update of
\begin{equation}
        p_{\mathrm{cut},t+\Delta t} = p_{\mathrm{cut},t}  \cdot e^{2 D_0 \Delta t}
        \label{eq:Dpp_cut}
\end{equation}
which we can then use to compute the fluxes between momentum bins according to Eq.~\ref{eq:n_flux_integral}-\ref{eq:e_flux_integral}.
Solving Eq.~\ref{eq:R_losses} with Eq.~\ref{eq:dpdt_Dpp} gives
\begin{equation}
    R_{D_\mathrm{pp}} = -2 D_0,
\end{equation}
as all prefactors cancel out.
We note that this approach is a simplification and neglects the stochastic component of the turbulent reacceleration. 
It does however counter the radiative loss processes of high-momentum electrons and with that provides the basis for studying the stabilization of radio haloes in galaxy clusters against the cooling times of their electrons.
The exponential dependency on $D_0$ and the timestep $\Delta t$ puts a significant timestep constraint on the simulation and needs to be handled with care.
We therefore subcycle the solver if $D_0 \Delta t > 10^{-5}$.
This increases the runtime by roughly a factor of 2-3, which would make it unfeasible in the case of the full cosmological box.
We therefore switched this mode off in the box and ran comparison zoom-in simulations with the mode switched on.

\section{Magnetic field models}
\begin{figure}[h]
    \centering
    \includegraphics[width=8cm]{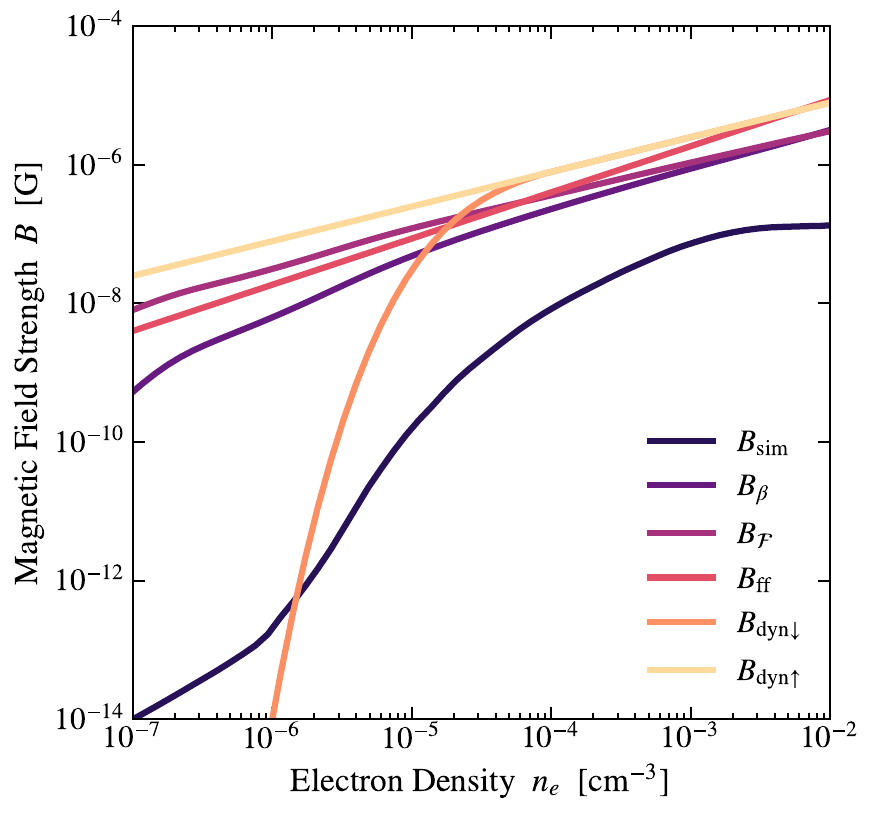}
    \caption{Scaling of our magnetic field models with electron density. For \bsim, \bbeta~and \bturb~we plot the mean magnetic field strength per density bin for the \crbox~simulation.}
    \label{fig:Bmodels}
\end{figure}
In Fig.~\ref{fig:Bmodels} we show the scaling of our magnetic field models with electron density.
We compute the scaling as the mean magnetic field strength per bin, obtained from the \crbox~simulation.

\section{Filament selection}
\begin{figure}[h]
    \centering
    \includegraphics[width=8cm]{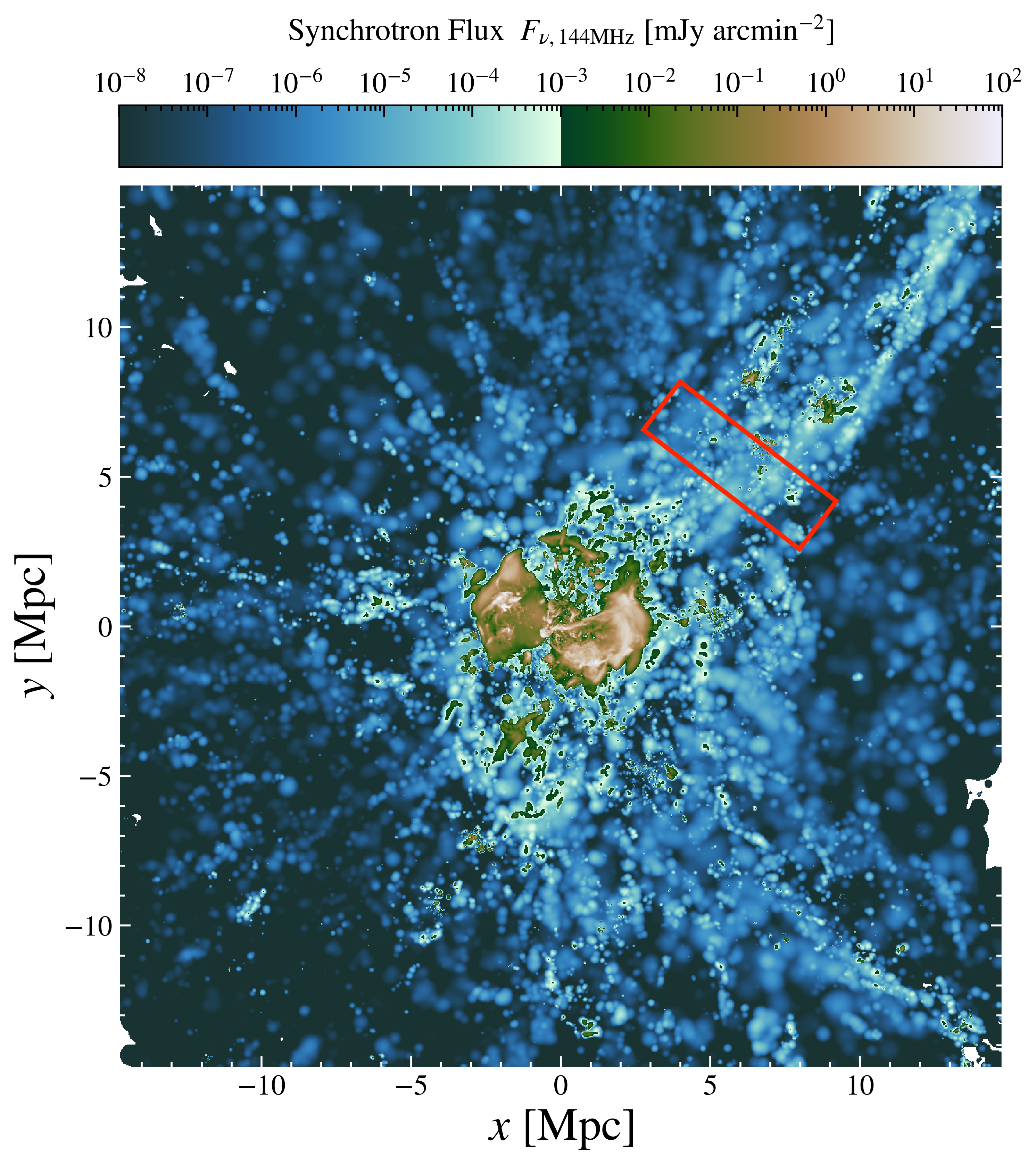}
    \caption{In red we sketch the region selected for the filament spectra in Fig.~\ref{fig:synch_spectrum}.
    For the background image, we chose the \comazoom~simulation with the \bdynh~magnetic field model.
}
    \label{fig:spectra_selection}
\end{figure}
In Fig.~\ref{fig:spectra_selection}, we sketch the selected region for the synchrotron spectra in Fig.~\ref{fig:synch_spectrum}. We note that this is a simplified illustration where we omit projection effects, as the filament does not lie perfectly in the xy plane of this projection.

%%%%%%%%%%%%%%%%%%%%%%%%%%%%%%%%%%%%%%%%%%%%%%%%%%

\end{document}